\documentclass{emulatemyapj}
\pdfoutput=1
\usepackage{amsmath,wasysym,pifont}
\usepackage{CJK}
\usepackage[colorlinks=true,breaklinks=true,citecolor=blue,linkcolor=blue,urlcolor=blue]{hyperref}

\begin{document}

\shorttitle{Photometry's bright future}
\shortauthors{M. Hippke \& D. Angerhausen}
\title{Photometry's bright future: \\Detecting Solar System analogues with future space telescopes}

\author{Michael Hippke}
\email{hippke@ifda.eu}
\affil{Institute for Data Analysis, Luiter Stra{\ss}e 21b, 47506 Neukirchen-Vluyn, Germany}

\author{Daniel Angerhausen}
\email{daniel.angerhausen@nasa.gov}
\affil{NASA Postdoctoral Program Fellow, NASA Goddard Space Flight Center, Greenbelt, MD 20771, USA}

\begin{abstract}
Time-series transit photometry from the \textit{Kepler} space telescope has allowed for the discovery of thousands of exoplanets. We explore the potential of yet improved future missions such as \textit{PLATO 2.0} in detecting solar system analogues. We use real-world solar data and end-to-end simulations to explore the stellar and instrumental noise properties. By injecting and retrieving planets, rings and moons of our own solar system, we show that the discovery of Venus- and Earth-analogues transiting G-dwarfs like our Sun is feasible at high S/N after collecting 6yrs of data, but Mars and Mercury will be difficult to detect due to stellar noise. In the best cases, Saturn's rings and Jupiter's moons will be detectable even in single transit observations. Through the high number ($>$1 bn) of observed stars by \textit{PLATO 2.0}, it will become possible to detect thousands of single-transit events by cold gas giants, analogue to our Jupiter, Saturn, Uranus and Neptune. Our own solar system aside, we also show, through signal injection and retrieval, that \textit{PLATO 2.0}-class photometry will allow for the secure detection of exomoons transiting quiet M-dwarfs. This is the first study analyzing in-depth the potential of future missions, and the ultimate limits of photometry, using realistic case examples.

\end{abstract}


\section{Introduction}
High precision, high duty time-series photometry from \textit{Kepler} has contributed to numerous fundamentally new discoveries in the exoplanet field (e.g. \citet{Borucki2010,Burke2014}). The primary mission ended in 2014 after finding thousands of planets and candidates \citep{Mullally2015}, with the technical failure of two reaction wheels. The extended \textit{K2} mission \citep{Howell2014} continues to deliver data and new planets \citep{Vanderburg2015,Foreman2015}. After this huge success, the next spacecraft photometry missions with improved technology are expected for 2017 (Transiting Exoplanet Survey Satellite, \textit{TESS} \citep{Ricker2014}) and 2024 (Planetary Transits and Oscillations of stars, \textit{PLATO 2.0} \citep{Rauer2014}). We asked the obvious question: What can we expect from these missions in the very best case? And ultimately: Assuming near-perfect photometric technology at some point in the future, what can we expect from photometry as such? Where are the fundamental limits? 

To begin, we will examine and define instrumental and stellar noise (section~\ref{sec:instruments}). Afterwards, we will discuss the \textit{Kepler}, \textit{TESS} and \textit{PLATO 2.0} mission designs and limitations (section~\ref{sec:sim}). We will not focus on the high numbers of discoveries, but on the very best cases with respect to their instrumental noise. In section~\ref{sec:inject}, we will present the view of a distant observer at \textit{our} planets transiting \textit{our} Sun, assuming near-perfect photometry. Inversely, this is what we can expect from future space missions when it comes to finding solar system analogues. 

We will conclude with an outlook to the limits of photometry.

\section{Examining the noise}
\label{sec:instruments}
Noise in data is often the limiting factor of data analysis. Noise in exoplanet transit photometry is caused by instrumental imperfections ($N_{I}$) and stellar jitter ($N_{S}$). In the following, we will discuss both parts separately. The total noise $N$, assuming Gaussian distribution, is then calculated as: 
\begin{equation}
   N=\sqrt{N_{I}^{2}+N_{S}^{2}}
\end{equation}

\subsection{Stellar noise characterization}
\label{sub:stellarnoise}
Stellar noise occurs with different characteristics on all time scales. First considerations for the \textit{Kepler} mission by \citet{Batalha2002} were theoretical, due to the lack of precise data for other stars. In our sun, there is the 11-year solar activity cycle \citep{Schwabe1843,Usoskin2009}, a phenomenon we also find on different time scales in other stars \citep{Garcia2010}. The solar rotation of $\sim$27 days \citep{Bartels1934,Beck2000} introduces noise from spots, first noted by Galileo Galilei in 1612 \citep{Scheiner2010}. It has also been argued that the solar activity is modulated by planetary gravitational and electromagnetic forces acting on the sun, namely those by Mercury, Venus, Earth and Jupiter \citep{Scafetta2013}.

In the following, we will ignore trends longer than a few days, and assume they can be removed using filters such as \citet{Savitzky1964} used by \citet{Gilliland2011} to analyze \textit{Kepler} noise, median filtering (e.g. \citet{Carter2012,TalOr2013}) or polynomial fitting (e.g. \citet{Santerne2014,Gautier2012}). Instead, we focus on the jitter on time scales of planetary transits, mostly 1--10 hours \citep{Koch2010}. This jitter originates mainly from stellar oscillation modes, granulation at the surface of the star, and rotational activity \citep{Andersen2015}.

\begin{figure}
\includegraphics[width=\linewidth]{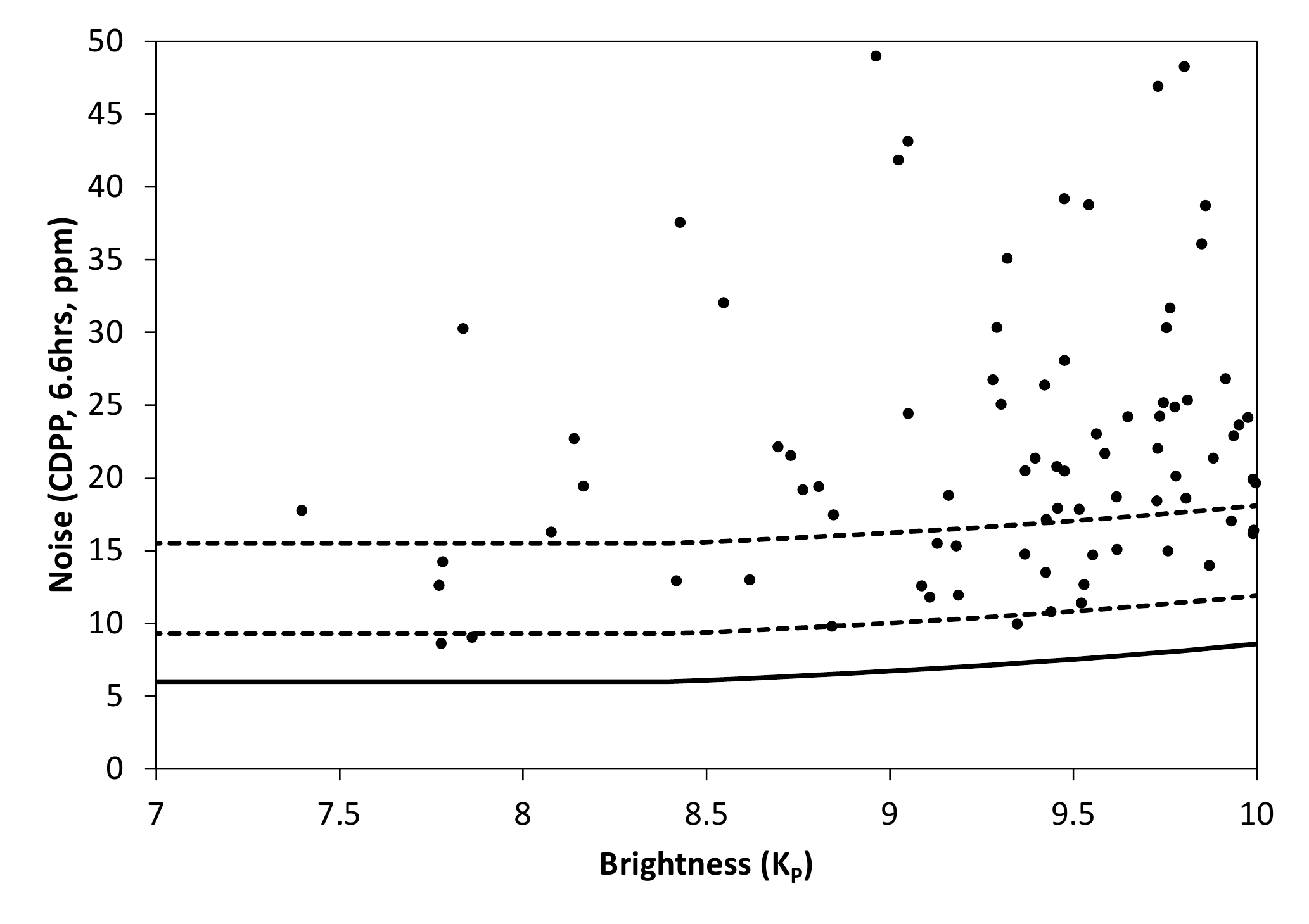}
\caption{\label{fig:jittersun}\textit{Kepler} instrumental noise (straight line) depends on the brightness of the star. This plot shows only the brightest end, and only G-dwarfs. Our sun would exhibit a total noise (instrumental plus stellar) in between the dashed lines, which give the limits for the active (upper) and quiet (lower) sun. Few other stars are more quiet than our sun, and many are more active.\\}
\end{figure}

Our sun's noise varies by a factor of $\sim$2 during the 11-year solar cycle, from 7.8ppm (2007.77, quiet period) in 6.5hrs bins to 14.7ppm (2002.39, active period) \citep{Gilliland2011,Frohlich1997}. This is at the quiet side of G-type stars, of which the most quiet 1\% have 6ppm, with a total cut-off at 5ppm \citep{Christiansen2012,Basri2013}. Although the noise measures and results differ slightly between these authors, it can also be seen from Figure~\ref{fig:jittersun} that there are few stars more quiet than our sun. There is a strong dependence of stellar noise to temperature: Cooler stars are usually more active, so that among M-dwarfs only very few are as quiet as our sun, and most are around $\sim$50ppm. However, there might be a few extremely quiet (1--4ppm) G-dwarfs \citep{Hall2007}, theorized to exhibit a time of almost no spots, as was the case for our Sun during the Maunder Minimum \citep{Maunder1912,Zolotova2015}. The detection of such a fortunate case, where mainly granulation (1ppm) contributes to stellar noise, would allow for extreme observations, given near-perfect technology. As we have not detected such a very-low noise star yet, we will instead concentrate on the known quiet end of G- and M-dwarfs.

\begin{figure}
\includegraphics[width=\linewidth]{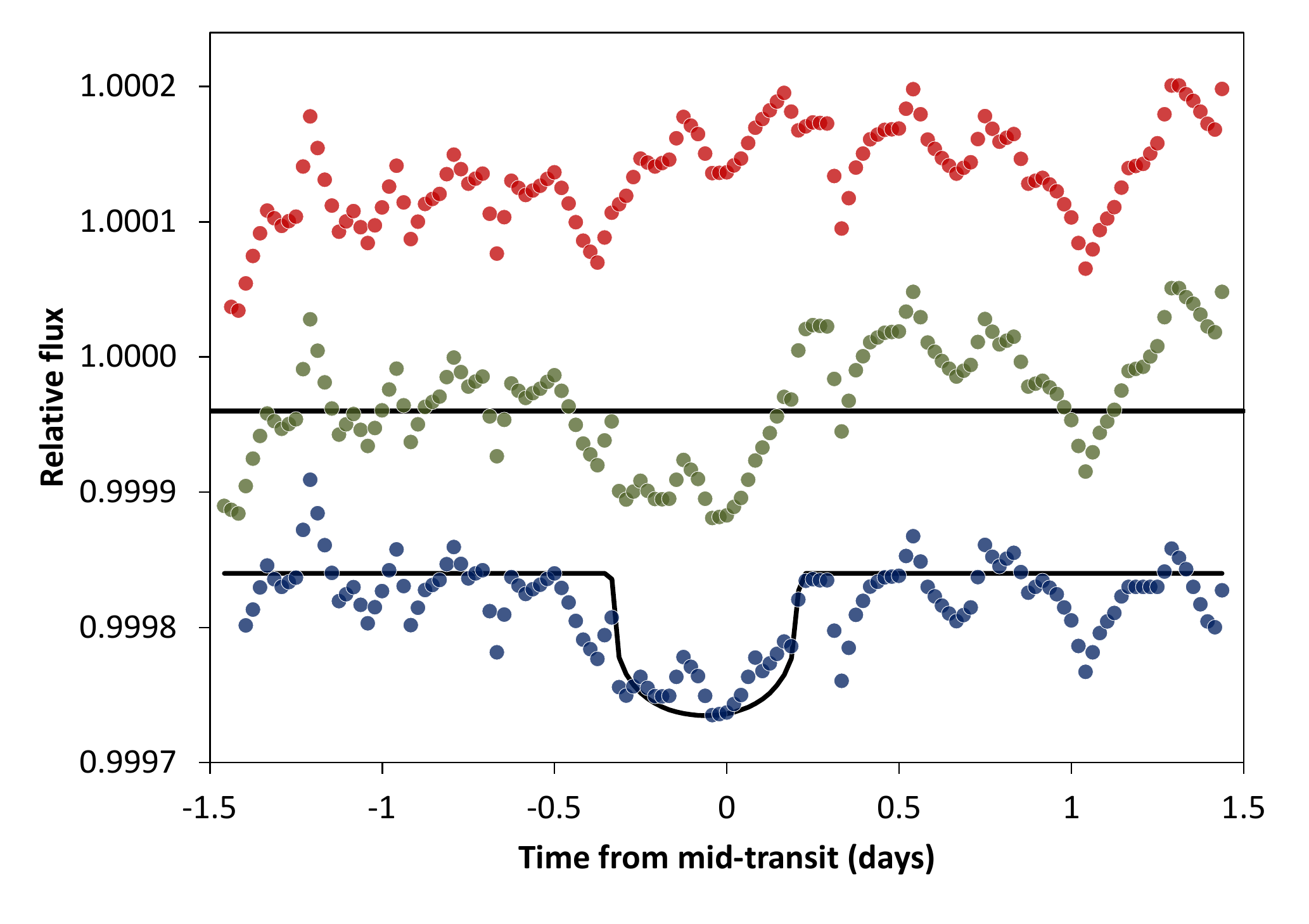}
\caption{\label{fig:sunjitter}Data for the sun from \textit{VIRGO/DIARAD} \citep{Frohlich1997} shows trends on time scales of hours and longer (top). We inject synthetic transit shapes into these raw data (middle). To retrieve the transit, we fit a long-time sliding median while masking the times of transit (bottom). Instrumental noise ($\sim$0.4ppm) is smaller than the symbol size.\\}
\end{figure}

Noise in sun-like stars, on this timescale, originates mostly from asteroseismic oscillations. Solar-like oscillations, mostly acoustic or pressure (p) modes, are reported with their highest amplitudes at frequencies between 29.9$\mu$Hz (10 hours) and 3619$\mu$Hz (4 minutes) among \textit{Kepler} stars \citep{Huber2012}. Regarding the solar noise properties on time scales of $\sim$hours, it can clearly be seen from Figure~\ref{fig:sunjitter} (top) that the jitter is not Gaussian. Clearly, there are trends, mostly from oscillations and spots, lasting a few hours, showing spots appearing and disappearing on the disc. This is the same timescale on which exoplanet transits occur, so that these trends cannot easily be filtered out. Time-correlated red (Brownian) noise becomes more and more Gaussian with stacks from different epochs, because the correlation of the total noise decreases (\citet{Barnes1966} and references therein). This will be explained in more detail in section~\ref{sub:red}.

Another noise source is contamination of the starlight from background (or foreground) stars in the aperture, which can occur when the angular separation of the contaminator is smaller than the spatial resolution of the instrument. For \textit{Kepler}'s (low) spatial resolution, this is a problem in some cases, but not in general: ``The overall increase of median and mean noise (...) are only 0.2 and 0.1 ppm'' \citep{Gilliland2011}.

\subsection{Stellar noise modeling}
\label{sub:noisemodel}
\begin{figure}
\includegraphics[width=\linewidth]{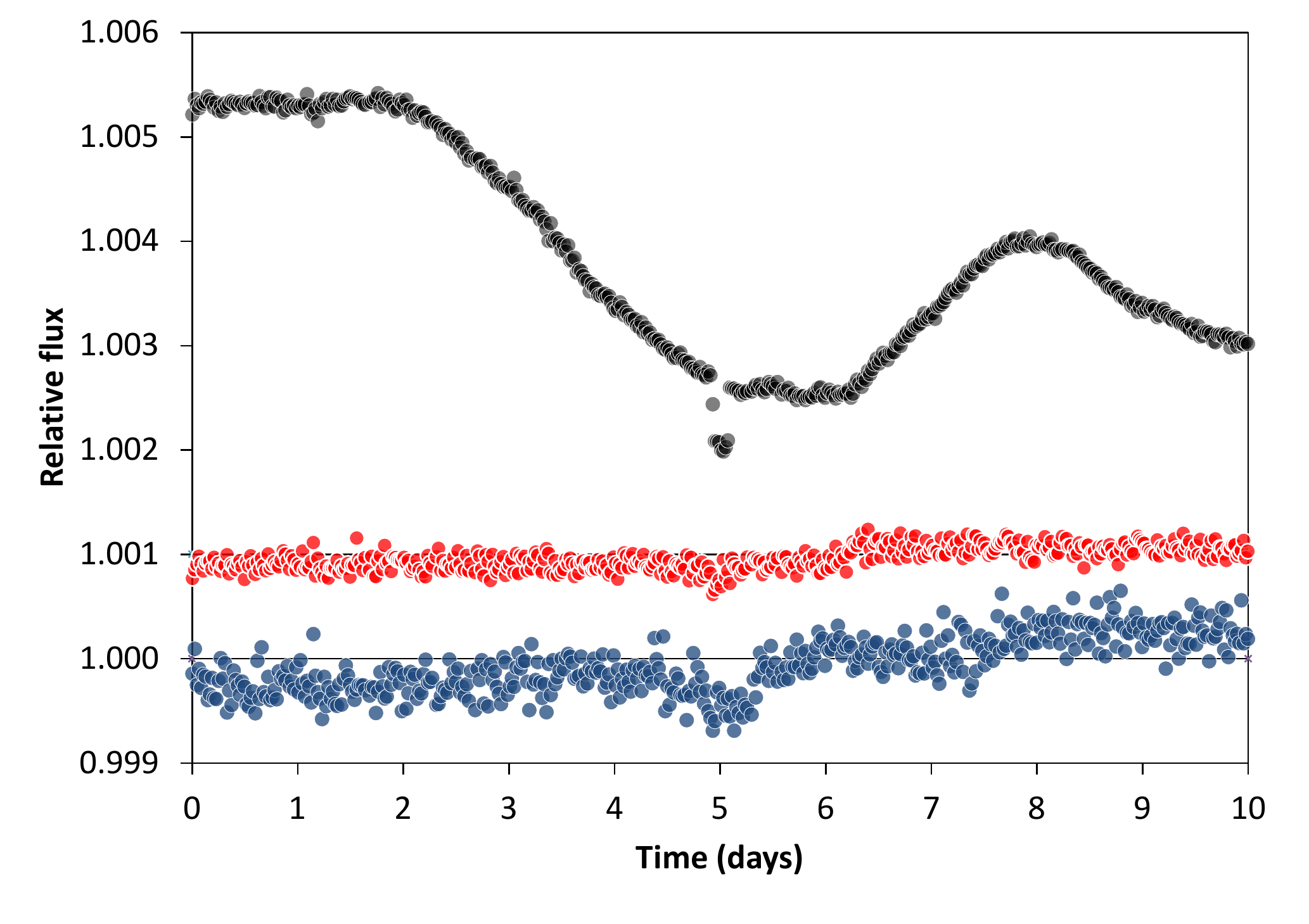}
\caption{\label{fig:stellartrends}After removing instrumental trends, different characteristics in stellar noise are evident. Kepler-96 (top, black) shows strong long-time (weeks) variations, but is as quiet as the other examples on hours to days timescale. Kepler-197 (middle, red) is a star with low noise on all time scales, like our Sun. Kepler-264b (bottom, blue) is as quiet long-time, but shows more prominent short-time (hours, days) trends. All stars have been centered to exhibit a transiting planet at T=5d.\\}
\end{figure}

Stars show brightness variations on different time scales, with different amplitudes, and characteristics. On the extreme side, there are for example RRab Lyrae with an amplitude of $\sim$50\% over 0.5d, but near-perfect repetition in some cases \citep{Smith2004,Szabo2014}. On the quiet side, there are stars like our Sun, or Kepler-197 \citep{Rowe2014}, which show very low (0.1\%) long-time (months) variation, but stochastic trends on an hour to day timescale. Intermediate cases are most common, e.g. Kepler-264b \citep{Rowe2014,Hippke2015} with strong short-time trends, or Kepler-96 \citep{Marcy2014} with prominent long-time trends. We show these examples in Figure~\ref{fig:stellartrends} and use them for the following discussion of suitable noise models.

The most fortunate case is to have trends that are on a much longer timescale than the transit signature. Kepler-96 is an example for this behavior, where a ``stellar noise model'' using splines or polynomials will successfully remove the trends which are induced mostly by stellar rotation. The short transit times can be masked for the detrending, and the transit analysis can be done afterwards. A simultaneous fit would only complicate the process, and have no additional benefit. The limits of this method are reached in cases where the trends are less sinusoidal, as is the case for CoRoT-7b \citep{Haywood2014,Barros2014}. Then, Gaussian Processes (GPs) are suitable to treat these systematics. Their main advantage is to naturally handle correlations irrespective of their origin, by specifying high-level properties of the covariance \citep{Evans2015}. Commonly used correlations are radial-velocity measurements \citep{Haywood2014}, different wavelengths \citep{Evans2015}, or instrumental systematics such as the drift in the roll-angle of \textit{Kepler's} K2 mission \citep{Aigrain2015,Foreman2015}.

When the noise timescale gets closer to the transit duration, other methods must be chosen. Most prominently, the wavelet-based formalism described in \citet{Carter2009} is being used, e.g. by \citet{Huber2013b} in their analysis of spin-orbit misalignment in the multiplanet system Kepler-56, or by \citet{Barclay2015} in their confirmation of Kepler-91 being a ``giant planet orbiting a giant star''. Using wavelets, the noise is modeled by the sum of two stationary (Gaussian) processes; one is uncorrelated in time, and the other has a spectral power density as $1/f^\gamma$. 

In any such model, parameters must be estimated and validated individually for each star, using Monte-Carlo simulations. This is also true for competing methods, such as the ``time-averaging'' method \citep{Pont2006}, or the ``residual-permutation'' method (e.g., \citet{Jenkins2002}). Consequently, stellar noise modeling is not applied to the large number (thousands) of planets found with automatic pipelines. For interesting individual cases, however, stellar noise modeling can and should be used. It is important to note that these cases have, beforehand, been detected \textit{without} modeling, so that any detection must pass this initial threshold\footnote{Further advantages in computer algorithms might introduce methods that can self-adjust to individual cases and validate the result.}. In the course of this paper, we choose the same approach: We use the ``standard'' pipeline without noise modeling, but introduce it after the detection of the Earth, which can be assumed to be of distinguished interest for any distant observer. Noise modeling reaches its current limits for signals so small that they become indistinguishable from permanent noise features; for our Sun, this would be the case for signals on ppm-level as for Earth's moon. Therefore, we limit our example to Earth itself.

\subsection{Instrumental noise}
\label{sub:instrumentalnoise}
Even the most perfect instrument will produce \textit{some} noise. Fundamentally, this originates from the fact that photons (starlight) and electrons (detector) are quantized \citep{Einstein1905}, so that only a finite number can be counted in a given time. This phenomenon is the \textit{shot noise} \citep{Schottky1918}, which is correlated mostly to the brightness of the target. In addition, noise occurs from the readout of the CCD, when the small signal gets amplified. On a timescale of 6.5hrs (13 bins of $\sim$30min), the \textit{Kepler} instrumental noise for a $K_{P}$=12 star from Poisson (shot) and readout is 16.8ppm. Two other instrumental noise sources have been quantified for the Kepler spacecraft: Intrinsic detector variations (10.8ppm), and a quarter-dependent term (7.8ppm) \citep{Gilliland2011}. The total instrumental noise is then 20.4ppm. For reference, the design of the spacecraft expected instrumental noise of 17.0ppm.

\begin{figure*}
\includegraphics[width=0.5\linewidth]{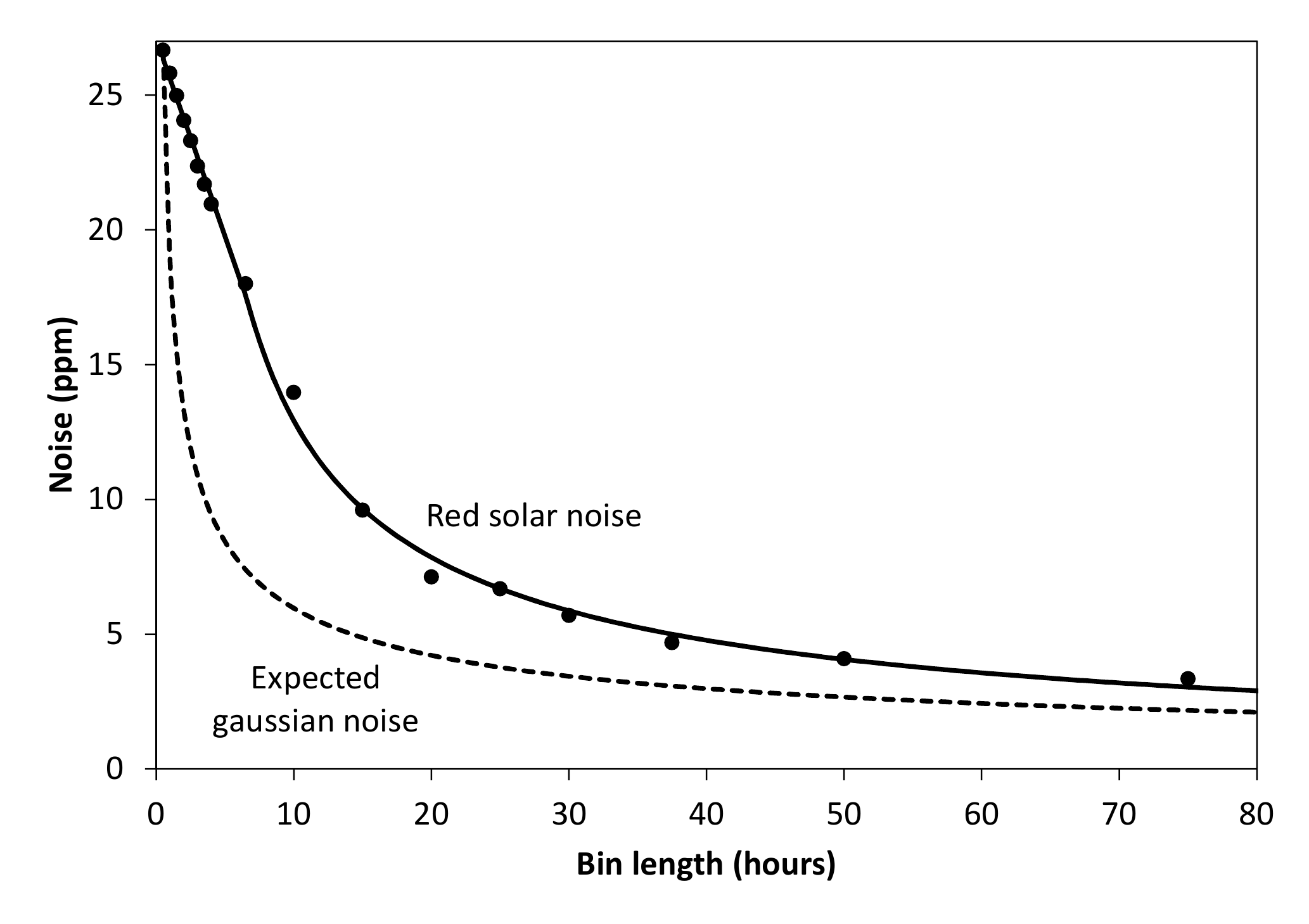}
\includegraphics[width=0.5\linewidth]{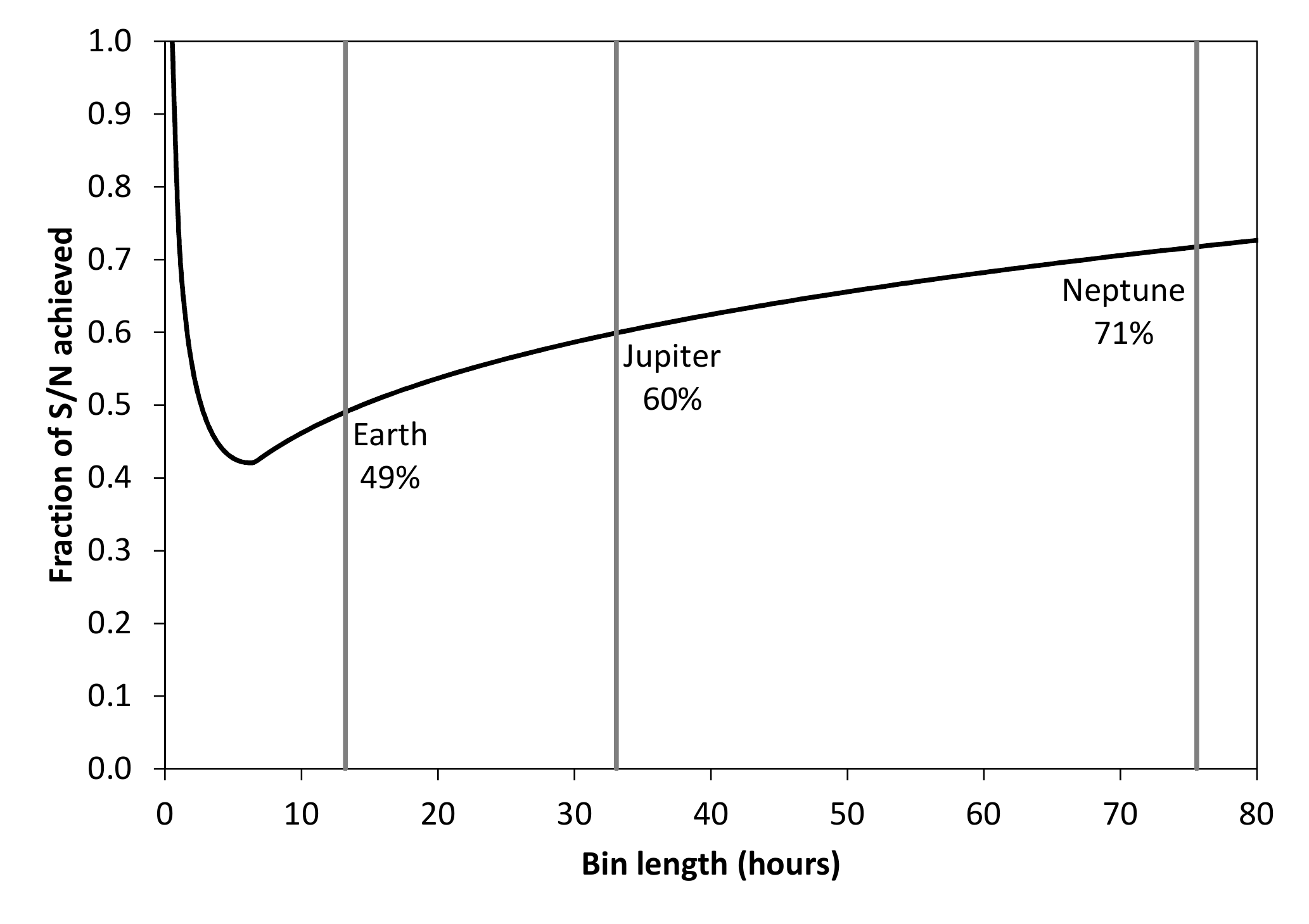}
\caption{\label{fig:snratio}S/N ratio for transits that can be achieved with real red-noise solar data, when compared to theoretical pure white Gaussian noise. Left: Observed (data points, straight line) and simulated Gaussian noise (dashed line). Right: O-C for the noise. The structural break at 7hrs is evident. For Earth-long transits, the achieved S/N is 49\%, and higher for longer transit durations.}
\end{figure*}

\subsection{Signal-to-noise definition and noise characterization}
\label{sub:red}
It is a common practice in exoplanet science \citep{Jenkins2002,Rowe2014} to define the signal-to-noise ratio as the depth of the transit model, compared to the out-of-transit noise:
\begin{equation}
{\rm S/N} = \sqrt{N_{T}}\frac{T_{dep}}{\sigma_{OT}}
\end{equation}
with $N_{T}$ as the number of transit observations, $T_{dep}$ the transit depth and $\sigma_{OT}$ the standard deviation of out-of-transit observations. This measure overestimates S/N for large planet-to-star radius ratios, and large impact parameters, but these configurations will be neglected in the present work. As pointed out by \citet{Fressin2013}, the detection of KOIs becomes unreliable for a S/N $\lesssim$ 10. This assumes Gaussian noise -- the noise distribution of our Sun, even during the quiet year 2007, is distinctly not Gaussian. It has strong time-correlated (red) noise features and a ``long tail''. Two statistical methods should be employed to quantify this. First of all, we recommend a skewness and Kurtosis tests for normality, e.g. by \citet{Shapiro1965}. For our sun, as expected, the difference to Gaussian noise is significant, at the 0.1\% level. After establishing that the noise is not Gaussian, we propose to measure the ``redness'' per bin length $L$, as described by \citet{Steves2010}:
\begin{equation}
   \sigma_{binned}=\sigma_{unbinned}^{L^{\beta}}
\end{equation}
For completely uncorrelated noise, $\beta=-1/2$, while for Gaussian noise $\beta=0$. Instead of giving $\beta$-values for each bin length, we chose to calculate the achieved percentage (where Gaussian noise is 100\%), and show the result in Figure~\ref{fig:snratio}. This gives the recoverable S/N of transits, when compared to Gaussian noise. As shown in the figure, the S/N in solar red noise is significantly lower. This issue originates from the simple fact that subsequent data points are time-correlated, and are thus more likely to have (nearly) \textit{the same error} (positive or negative) when binned together on the time axis. In other words: Red noise data doesn't bin as well as white noise data. The penalty of this noise characteristic is also time-correlated, because, over time, it diminishes. For an Earth-analogue transit (13.2hrs), only 49\% of the expected S/N can be achieved; 60\% for a Jupiter (33hrs) and 71\% for a Neptune (76hrs) (Figure~\ref{fig:snratio}). When stacking different years (of similar noise), the penalty is zero, as there is no time-correlation any more. The original intra-transit penalty cannot be recovered, of course, but stacking different epochs brings the expected (like Gaussian) stacking bonus.

As shown in Figure~\ref{fig:snratio}, the expected power law when binning data (for mainly Gaussian noise) begins at $\sim$7hrs. For shorter times, the best-fit is given by a linear correlation ($R^2=0.99$; best-fit power law gives lower $R^2=0.87$). The location of the structural break can be calculated using the test by \citet{Chow1960}, asking the question whether the coefficients in two regressions for different data sets are equal. The result is a clear structural break at the 0.1\% level, with a best-fit location at 7$\pm1$hrs. To sum up, our sun produces Gaussian noise on long ($\gg$1d) timescales, but suffers red noise punishment on shorter timescales.

\begin{figure*}
\includegraphics[width=0.5\linewidth]{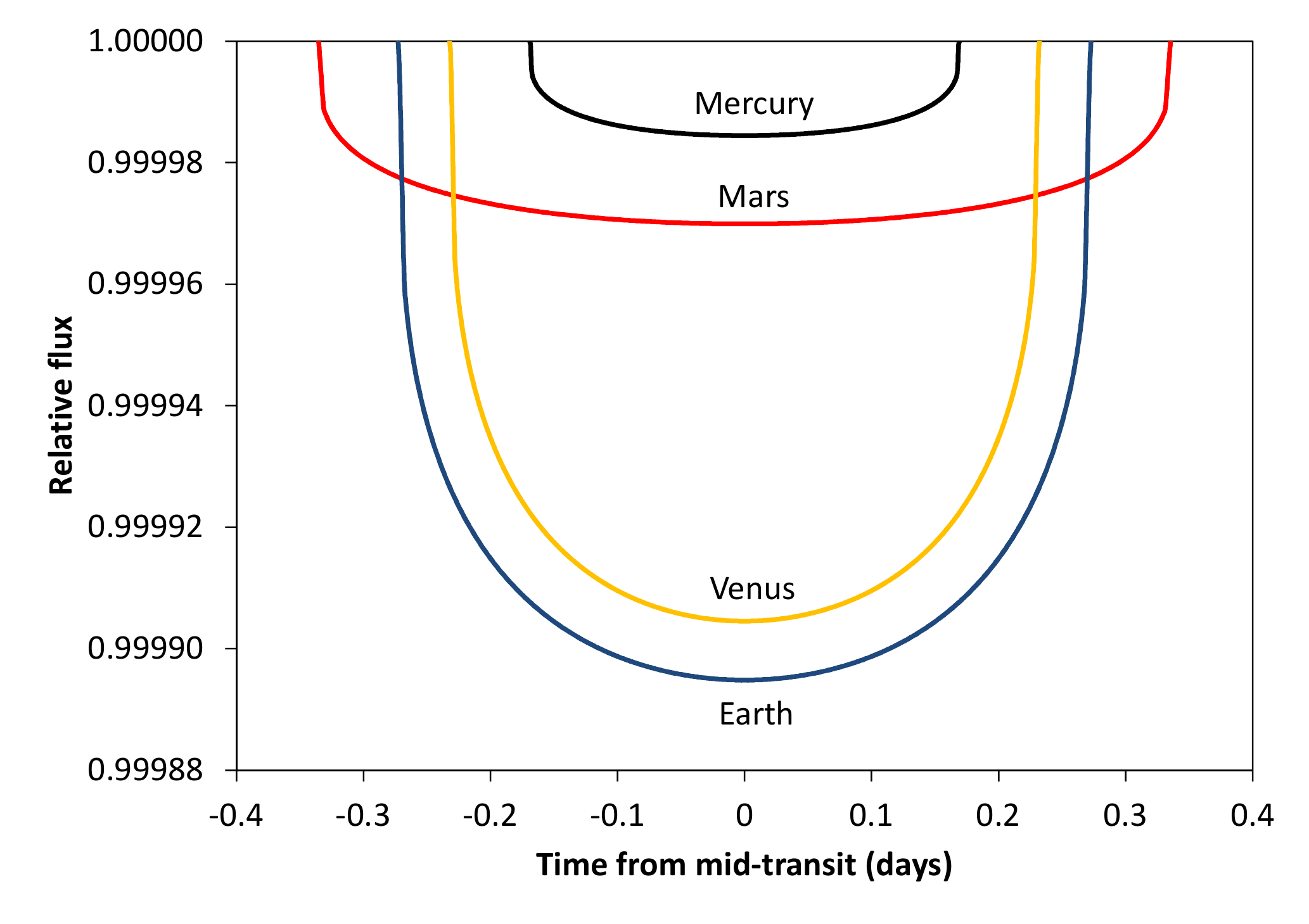}
\includegraphics[width=0.5\linewidth]{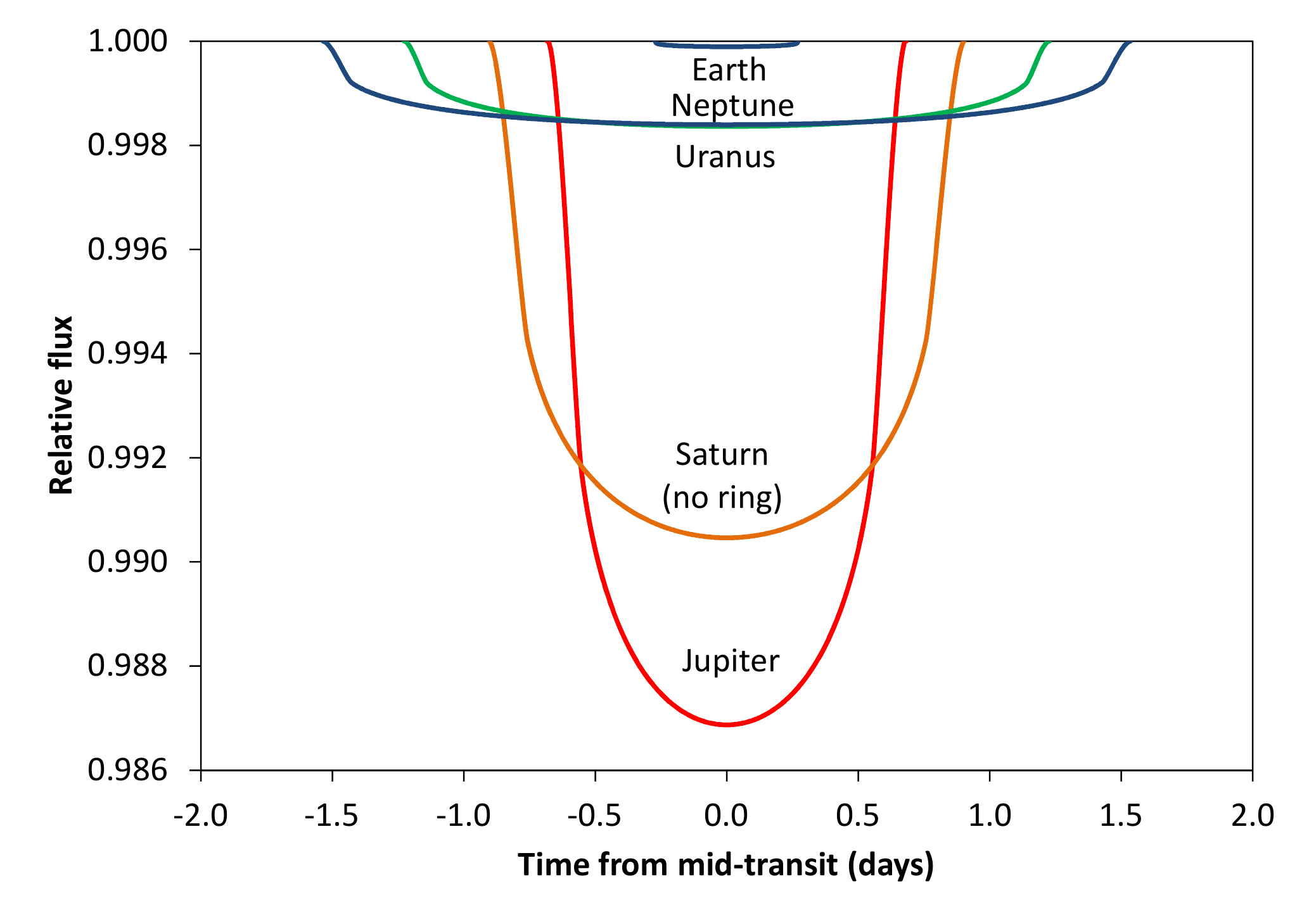}
\caption{\label{fig:planets}Transit light curves for solar system inner (left) and outer planets (right, with Earth for comparison). Impact factor has been set to 0, so that the transit duration is maximized. These artificial curves are used for the injections.}
\end{figure*}

\section{Method: Preparing the telescopes}
\label{sec:sim}
In this section, we will discuss the instrumental performance of the future \textit{TESS} and \textit{PLATO 2.0} missions. We will define the properties for two ``telescopes'' to use: The best-case \textit{PLATO 2.0} performance, and a theoretical near-perfect instrument, dubbed \textit{PERFECT}. These instruments will then be used in section~\ref{sec:inject} to inject and retrieve transits.

\subsection{Observations with \textit{TESS}}
The longest observing intervals for \textit{TESS} \citep{Ricker2014} during its 2-year mission will be one year (for parts of ths sky near the ecliptic pole). It will focus on the brightest ($I_{c}=4-13$) stars, suitable for follow-up observations with the \textit{James Webb Space Telescope} \citep{Gardner2006} (the areas of the sky with one year coverage from \textit{TESS} will also fall into the continuous viewing zone for \textit{JWST}) . As such, its mission design is not targeted at the very best photometric performance, although it is expected to be on par with \textit{Kepler} (\citet{Ricker2014}, their Figure 8). In this paper, we strive for best possible photometry, and are interested in solar system analogues, so that we will neglect \textit{TESS}, as it does ``not address the science case of characterizing rocky planets at intermediate orbital distances (a$>$0.3au, including the HZ) around solar-like stars, which remains unique for \textit{PLATO 2.0}.'' \citep{Rauer2014}.

\subsection{Observations with \textit{PLATO 2.0}}
Instrumental noise from \textit{PLATO 2.0} is expected to be as low as 10ppm on 30min timescale for bright (V=9) stars \citep{Zima2006}, or 8ppm in 1hr integrations \citep{Rauer2014}, when observed by many (up to 36) cameras simultaneously, thus producing $\sim$3ppm of instrumental noise on the same 6.5hrs timescale. While \textit{Kepler} observed $\sim$30 bright (V$<$8) stars, \textit{PLATO 2.0} will collect data for $\sim$3,000 -- giving a good chance for a useful share of quiet stars among them. Observing our sun through \textit{PLATO 2.0} would then give a stellar noise fraction of 74\% (quiet sun) to 84\% (active sun), making such an observation strongly limited by intrinsic stellar noise. For smaller, but generally more active M-dwarfs (with noise levels around 50ppm \citep{Basri2013}), this ratio can reach 95\% to 99\%, making the technology near-perfect and observations only limited by intrinsic stellar noise. This shows the ultimate limits of photometry as such, and the importance of understanding (and modeling) stellar noise. \textit{PLATO 2.0} has an expected lifetime of 6 years, giving a useful duration for multiple transits of Earth and Mars analogues.

For simulating realistic data as expected from \textit{PLATO 2.0}, an end-to-end simulator\footnote{\url{https://fys.kuleuven.be/ster/Software/PlatoSimulator}} evolved over the last decade \citep{Zima2006,MarcosArenal2014}. It takes into account effects down to the sub-pixel matrix, satellite orientation jitter, PSF convolution and all CCD-related noise sources. Using this simulator, we have created a flat lightcurve for a bright (V=9) star, resulting in Gaussian noise at $\sim$10ppm in 30min bins. These data are the instrumental basis of our \textit{PLATO 2.0} simulations in section~\ref{sec:inject}.

\subsection{The \textit{PERFECT} telescope}
 \textit{PLATO 2.0} is expected to launch in 2024 and end its nominal operation in 2030 \footnote{ \url{http://www2.warwick.ac.uk/fac/sci/physics/research/astro/plato-science/pre-launch/}}. For the following decade (the 2030s), one might imagine a successor mission, with even better instruments. We should assume that many noise sources can be reduced (or eliminated completely) using the knowledge and expertise gained throughout previous missions, such as pointing jitter and thermal variations. Sensor sensitivity, for instance, has improved over the decades and approaches $>$80\% quantum efficiency today, over the wavelength range 0.6--2.5$\mu$m (e.g., \citet{McGurk2014}.

For those stars where the photon flux (shot noise) is not the limiting factor, instrumental noise will be dominated by these other sources. As we have seen, total instrumental noise from \textit{Kepler} is at $\sim$20ppm, whereas we expect $\sim$10ppm from \textit{PLATO 2.0}. It will be interesting to check the perfect telescope with zero instrumental noise. As explained in section~\ref{sub:instrumentalnoise}, this is in principle unphysical, but a noise floor down to 0.4ppm as for \textit{VIRGO/DIARAD} might be achievable. We will keep (adopt) this 0.4ppm of instrumental noise for the virtual \textit{PERFECT} telescope. In section~\ref{sub:earth}, we will see that the difference between \textit{PLATO 2.0} and \textit{PERFECT} is negligible for the standard quiet G-dwarf, as stellar noise is the dominating noise source. For a very quiet M-dwarf, however, this can make an improvement of up to 50\% in S/N.

\begin{figure}
\includegraphics[width=\linewidth]{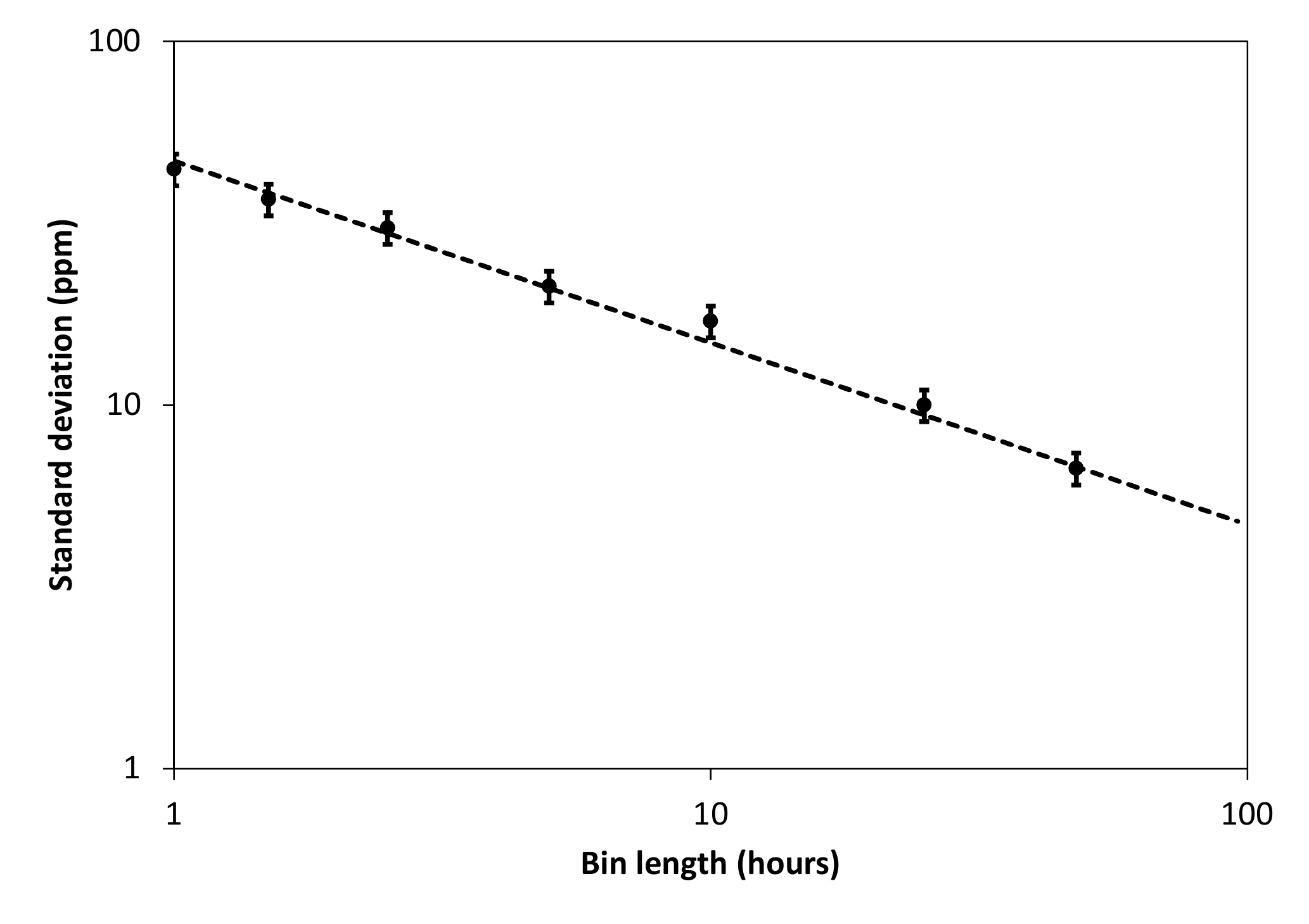}
\caption{\label{fig:mbin}Noise characterization of KIC7842386, the most quiet M-star out of 1015 (for which CDDP data is available) from \textit{Kepler}, with stellar noise CDDP=6.7ppm on 6.5hrs timescale. The dashed line represents (theoretical) pure Gaussian noise on a straight line in this log-log plot. Measured noise in bins (dots with error bars) is consistent with Gaussian noise.}
\end{figure}

\subsection{Target stars and planetary systems}
\label{sub:target}
As host stars, we choose a very quiet G2-dwarf like our sun -- in fact, we will simply use our sun's data from \textit{VIRGO/DIARAD} \citep{Frohlich1997,Appourchaux1997}. Its bandpass is comparable to the photometry space telescopes, and its instrumental noise is $<$0.4ppt and can thus be neglected. The performance is better than \textit{Kepler} or \textit{PLATO 2.0} due to the high flux the instrument can receive from our nearby Sun, when compared to stars many parsecs away. We have interpolated the data to 30min bins (analog to \textit{Kepler} long cadence data) and added  aforementioned $\sim$10ppm instrumental noise to these data in order to simulate a best-case observation by \textit{PLATO 2.0}. 

The other reference star is a 0.5$R_{\odot}$ M1-dwarf on the very quiet end of the distribution, exhibiting 7ppm of Gaussian (not time correlated) noise. From 1015 characterized \textit{Kepler} M-stars, two exhibit stellar (CDDP, \citet{Christiansen2012}) noise $<$7ppm. We have taken the most quiet one, KIC7842386 (6.7ppm on 6.5hrs timescale) and analyzed its noise characteristics. As can be seen in Figure~\ref{fig:mbin}, the noise is Gaussian within the errors. While the number of quiet M-stars might be low, we take the noise level of this fortunate example for our injections.

Due to their smaller radii, M1-dwarfs are particularly suitable to observe transits. Their absolute luminosity is smaller at $\sim3.5$\% of that of G-dwarfs, so that they need to be closer to the observer by a factor of $L_{\odot}/L_{*}^2$ to have the same apparent brightness for the same \textit{PLATO 2.0} instrumental noise.

Using these virtual instruments, we will observe different bodies transiting the G2-dwarf and the M1-dwarf. We inject planet transits following the standard \citet{Mandel2002} model, including quadratic limb-darkening with stellar metallicity \citep{Claret2011}, as implemented by \textsc{PyAstronomy}\footnote{\url{https://github.com/sczesla/PyAstronomy}}. For the G2-dwarf, we observe the planets of our own solar system, exhibiting a wide range of transit depths and -durations (Figure~\ref{fig:planets}). We set the impact parameter to zero, in order to maximize the transit duration. Special focus is on our Earth, including Earth's moon (section~\ref{sub:earth}), and the Jupiter system (section~\ref{sub:jupiter}). Finally, we will explore Saturn, including the flux gain caused by the forward-scattering of its rings (section~\ref{sub:saturn}), as well as Uranus and Neptune (section~\ref{sub:uranus}). To sum up, this section asks the question what we would see of \textit{our own} solar system if we were placed somewhere else in the galaxy, with near-perfect photometric equipment. 

For the M1-dwarf, we selected to test a 2.0$R_{\oplus}$ Super-Earth with Ganymede-sized (0.4$R_{\oplus}$) moon (section~\ref{sub:mdwarf}).

For all cases, we will use 30min integrations as the shortest bin available. This is comparable to \textit{Kepler}'s long-cadence (LC) bins, and guarantees sufficient sampling for all transits in this work. Only for Saturn's rings we find that a finer time resolution would be (marginally) beneficial.

\subsection{Injection and retrieval of transits}
We use the raw data from \textit{VIRGO/DIARAD} \citep{Frohlich1997} and inject the synthetic transit shapes as described in section~\ref{sub:target} and shown in Figure~\ref{fig:planets}. An exemplary step-by-step injection and retrieval is shown in Figure~\ref{fig:sunjitter}, here neglecting instrumental noise for clarity. To retrieve the transits, we apply a sliding median with boxcar length of $\sim$2 days, while masking the data points affected by the transit. For the longer transit durations of the outer planets, the boxcar length was increased accordingly, leaving slightly larger residuals. This method has the advantage that we can use solar data with negligible ($<$0.4ppm) instrumental noise, but fully preserving stellar noise. To these data, we add instrumental noise as described in section~\ref{sub:target}. This method allows us to control and modify noise sources separately.

\subsection{Transit probability}
The transit probability of any planet is low, so that many stars need to be observed, in order to collect a useful sample (Kepler observed $>$100,000 stars). For circular orbits, the transit probability can be calculated as \citep{Borucki1984}:
\begin{equation}
   P_{tr}=\frac{R_{*}}{a}
\end{equation}
where $R_{*}$ is the radius of the host star and $a$ is the semi-major axis of the planet's orbit. This gives transit probabilities e.g. for Earth as 0.47\%, and Jupiter 0.041\%: In order to potentially detect one Earth-analogue, one must on average survey 213 G-dwarfs that host an Earth like planet, and 2,439 that host a Jupiter. Assuming $\eta_{\oplus}=0.1$ \citep{Burke2015}, one must survey 2130 G-dwarfs to detect an Earth-analogue. Therefore, observing a large number of (dimmer and dimmer) stars is required at high sensitivity. The instrumental performance of \textit{PLATO 2.0} is expected to be sufficient for the detection of $\sim$150 Earth-sized planets on 365d-orbits (\citet{Rauer2014}, their Figure 5.5).

\section{Results: Observing transits with \textit{PLATO 2.0} and \textit{PERFECT}}
\label{sec:inject}

\begin{figure}
\includegraphics[width=\linewidth]{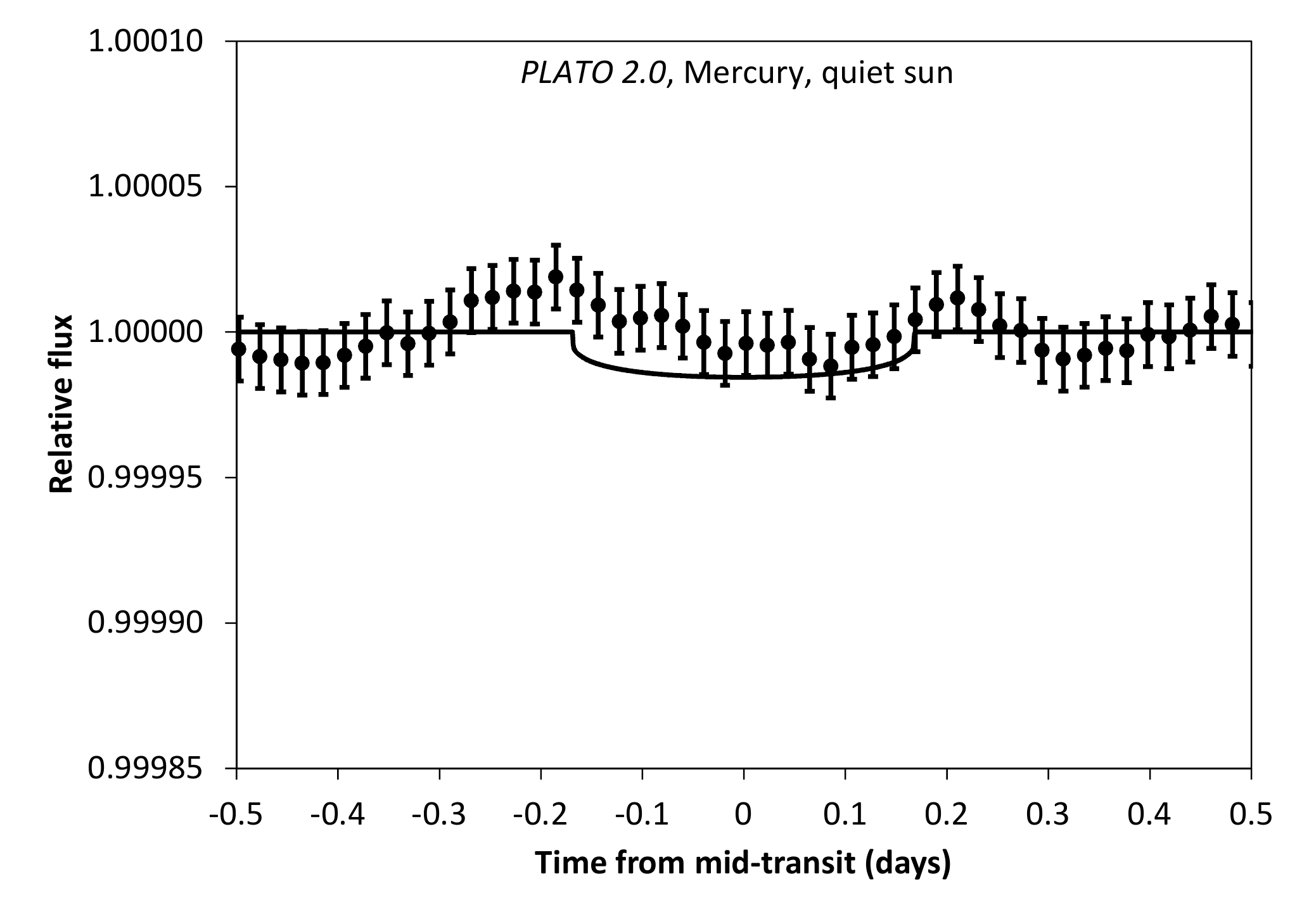}

\includegraphics[width=\linewidth]{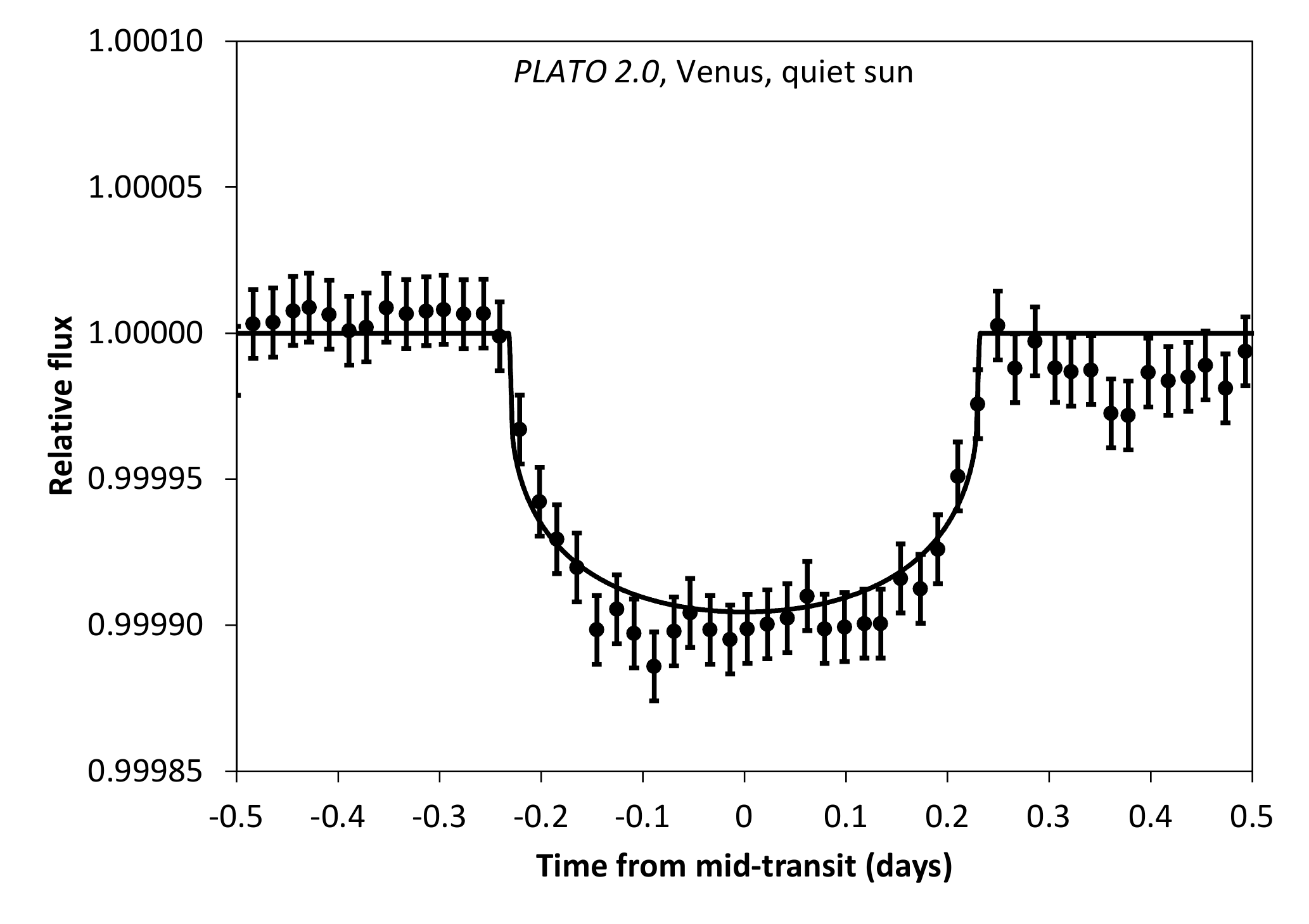}

\includegraphics[width=\linewidth]{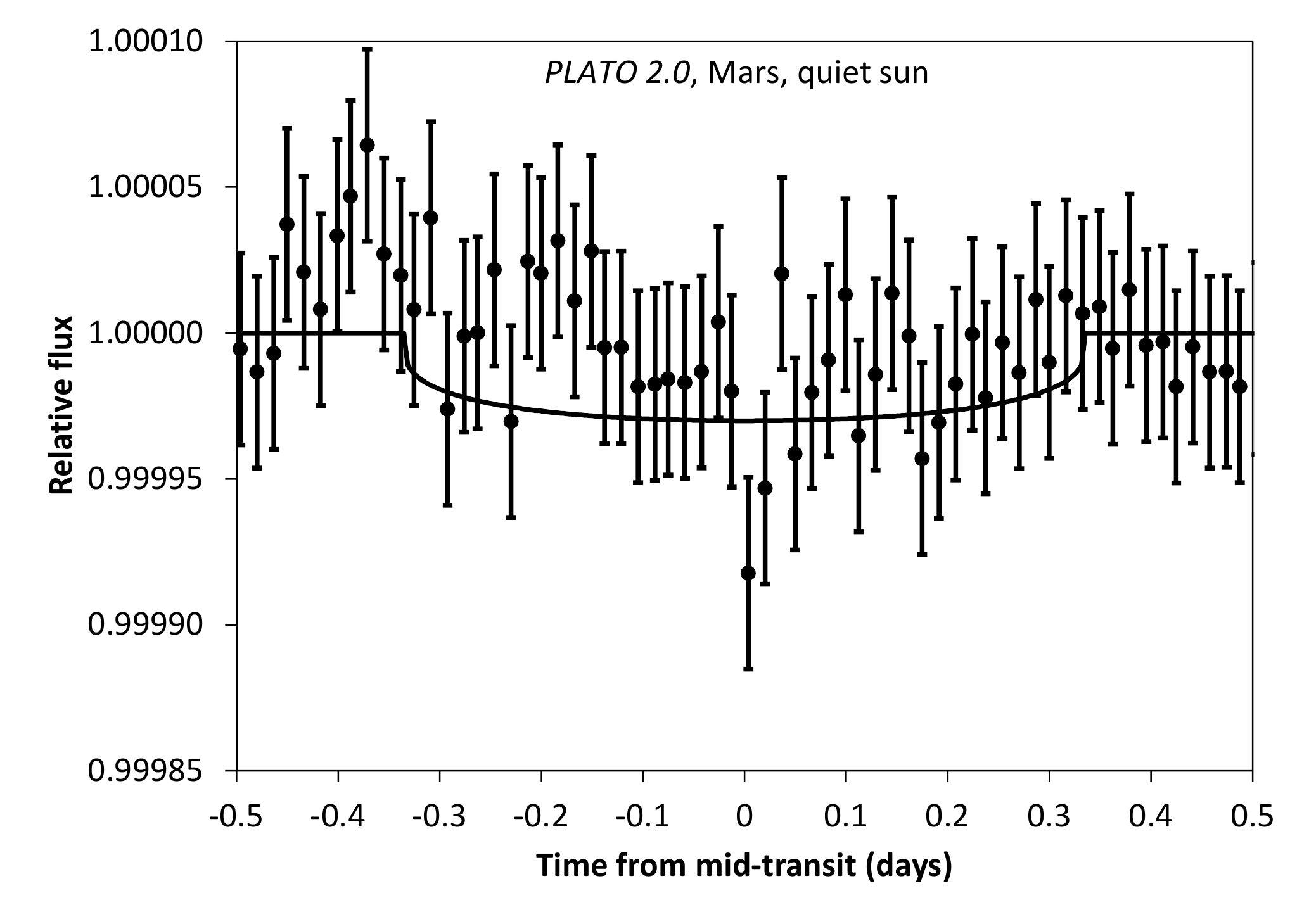}
\caption{\label{fig:mercury}Transits of Mercury, Venus, and Mars with six years of \textit{PLATO 2.0} coverage and a quiet sun. Due to the decreasing transit frequency, the error bars get larger in this order. Transits of Venus are clearly detected, but Mercury and Mars are not recoverable. For better comparison, all axes are identical and the same as for the Earth transits in Figure~\ref{fig:earth}.}
\end{figure}

\subsection{Mercury, Venus and Mars}
Mercury is the smallest (0.38$R_{\oplus}$) planet in the solar system with a transit depth of only 13.1ppm. On the other hand, its short period of 87.97 days allows for the greatest number of observed transits for a given observation time. When collecting six years of \textit{PLATO 2.0} data, and no data loss occurs during transit, a total of 25 transits can be recorded. We expect a nominal S/N=7.0 from this stack, but as can be seen in Figure~\ref{fig:mercury}, this is not sufficient for a detection due to the red noise characteristics. We conclude that Mercury analogues will likely not be found with photometry around normal G-dwarfs, when neglecting noise-modelling.

The next planet, Venus, is somewhat more interesting, as it is roughly (0.95$R_{\oplus}$, 80.6ppm) Earth-size, and transits more frequently due to its smaller orbit. On the other hand, its transit duration is also shorter than Earth's (0.46d vs. 0.55d), giving roughly similar detection S/N (Venus: 23.8; Earth: 28.6) for both when using six years of \textit{PLATO 2.0} observations. Venus orbits inside of the inner edge of the habitable zone \citep{Kopparapu2013}, but if orbiting a slightly less luminous stars, it might be habitable.

Mars detectability, then, suffers from smaller size (0.53$R_{\oplus}$, 25.4ppm) and longer orbit (only 3 transits in 6 years). Its transit duration (0.67d) helps only marginally with the transit detection, giving an insufficient S/N=7.2, comparable to Mercury.

\begin{figure*}
\includegraphics[width=0.5\linewidth]{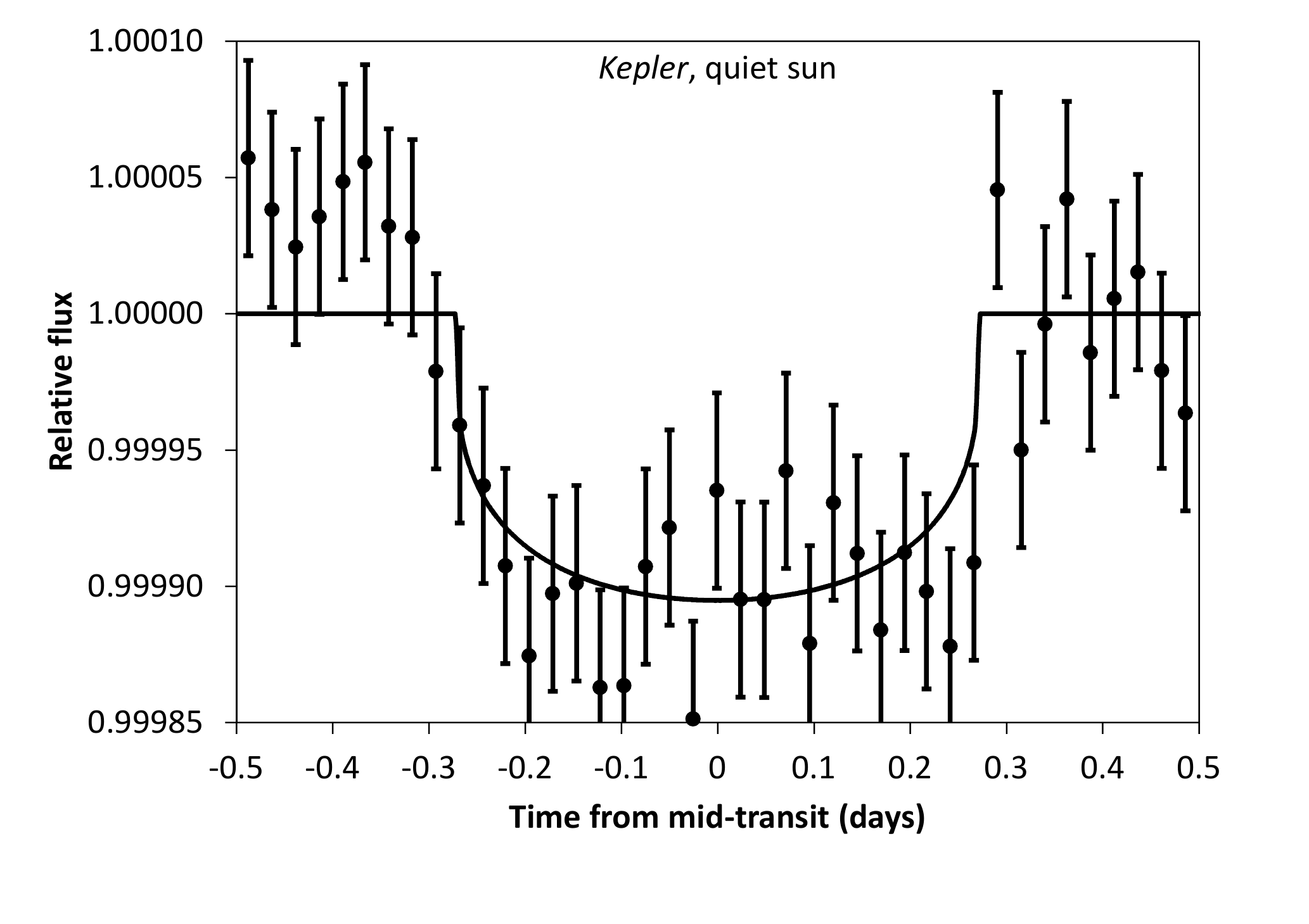}
\includegraphics[width=0.5\linewidth]{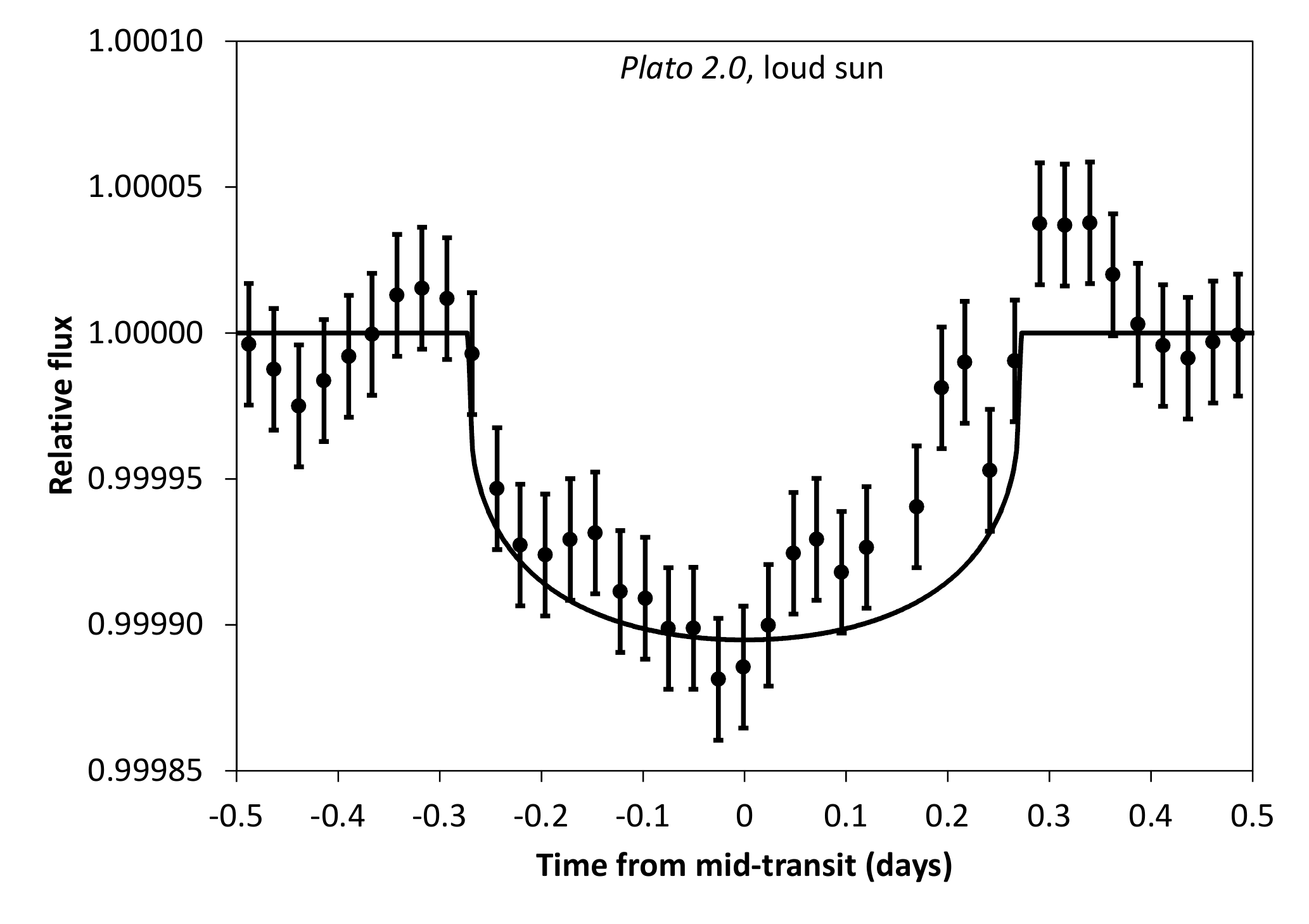}

\includegraphics[width=0.5\linewidth]{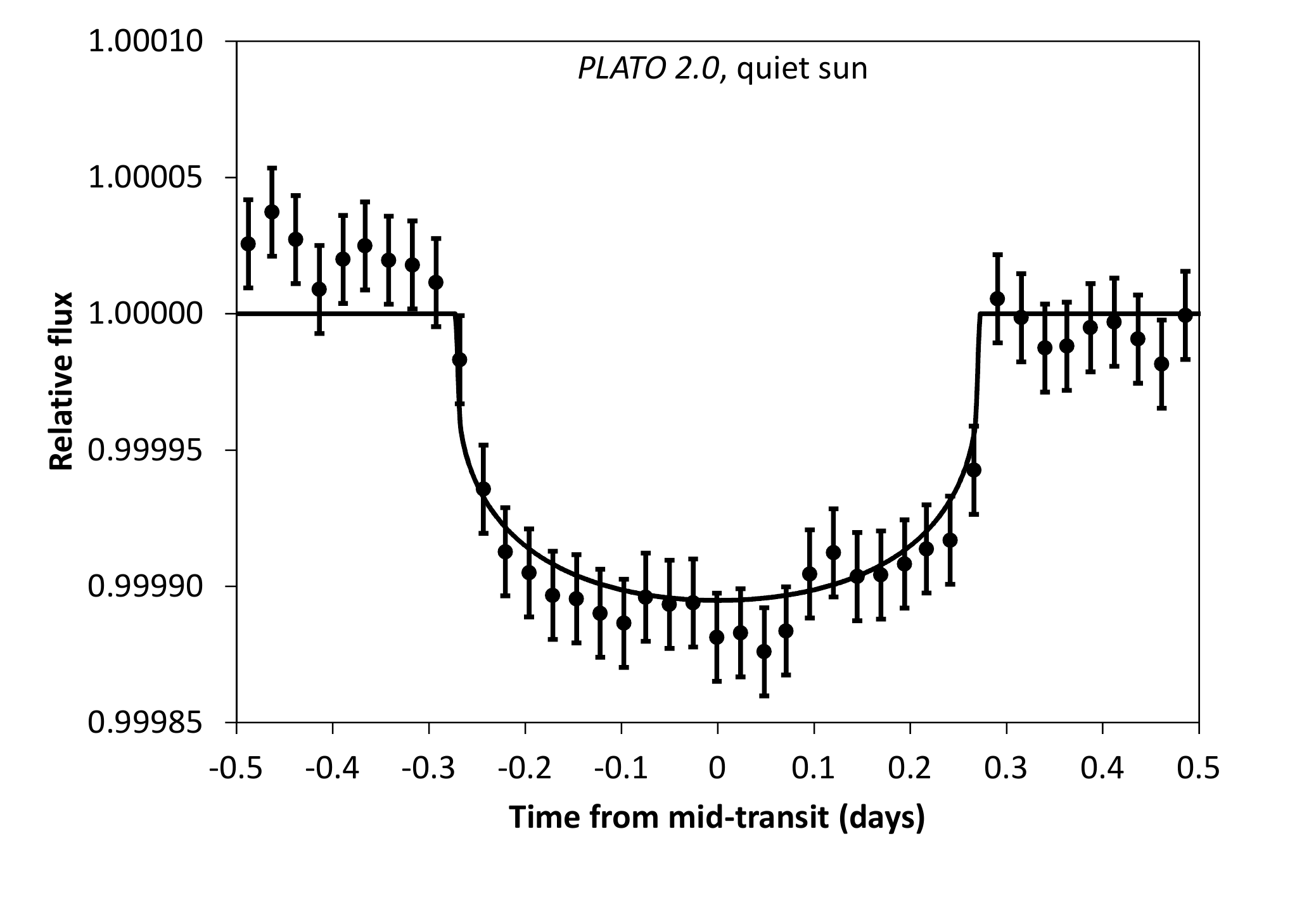}
\includegraphics[width=0.5\linewidth]{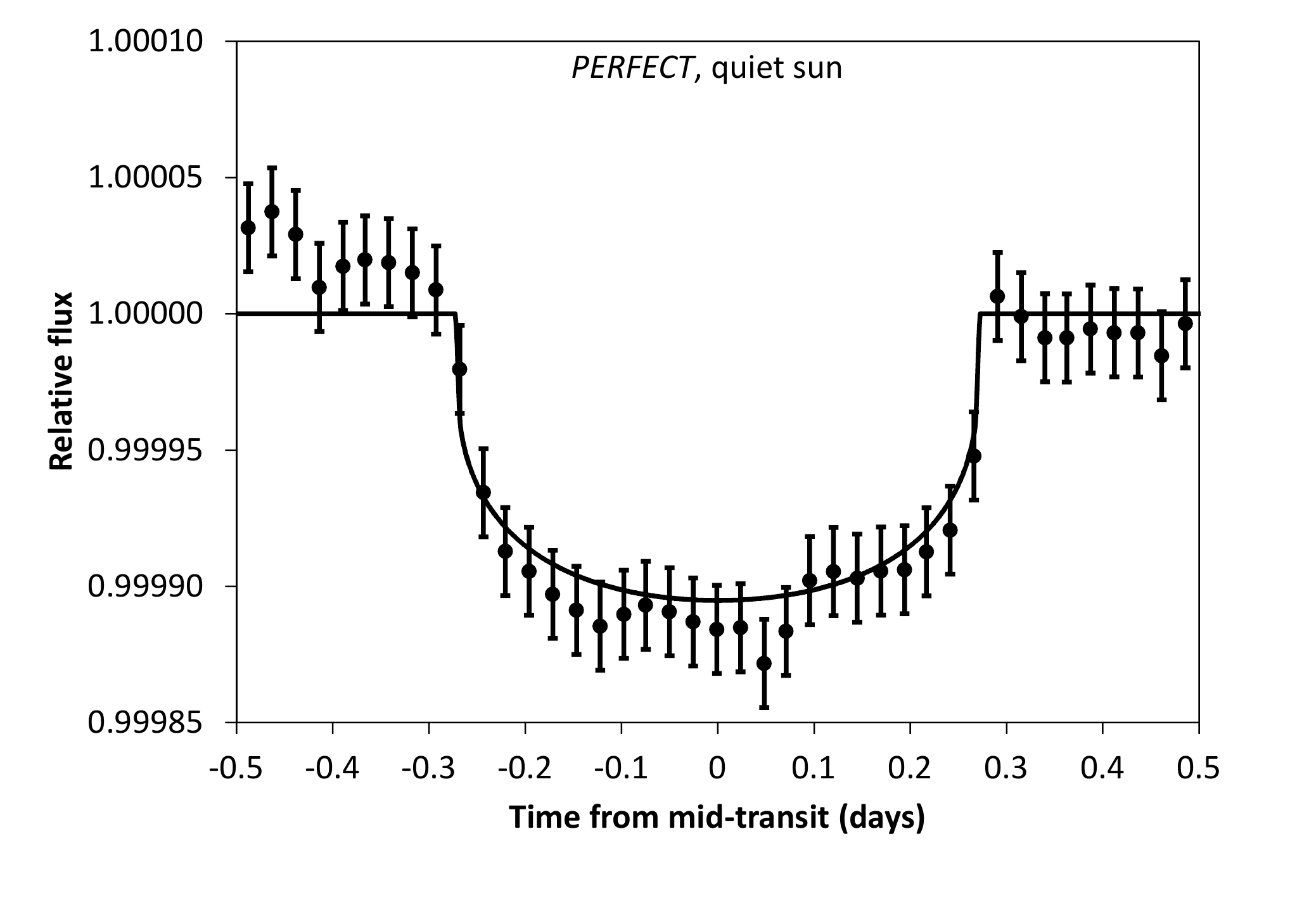}

\includegraphics[width=0.5\linewidth]{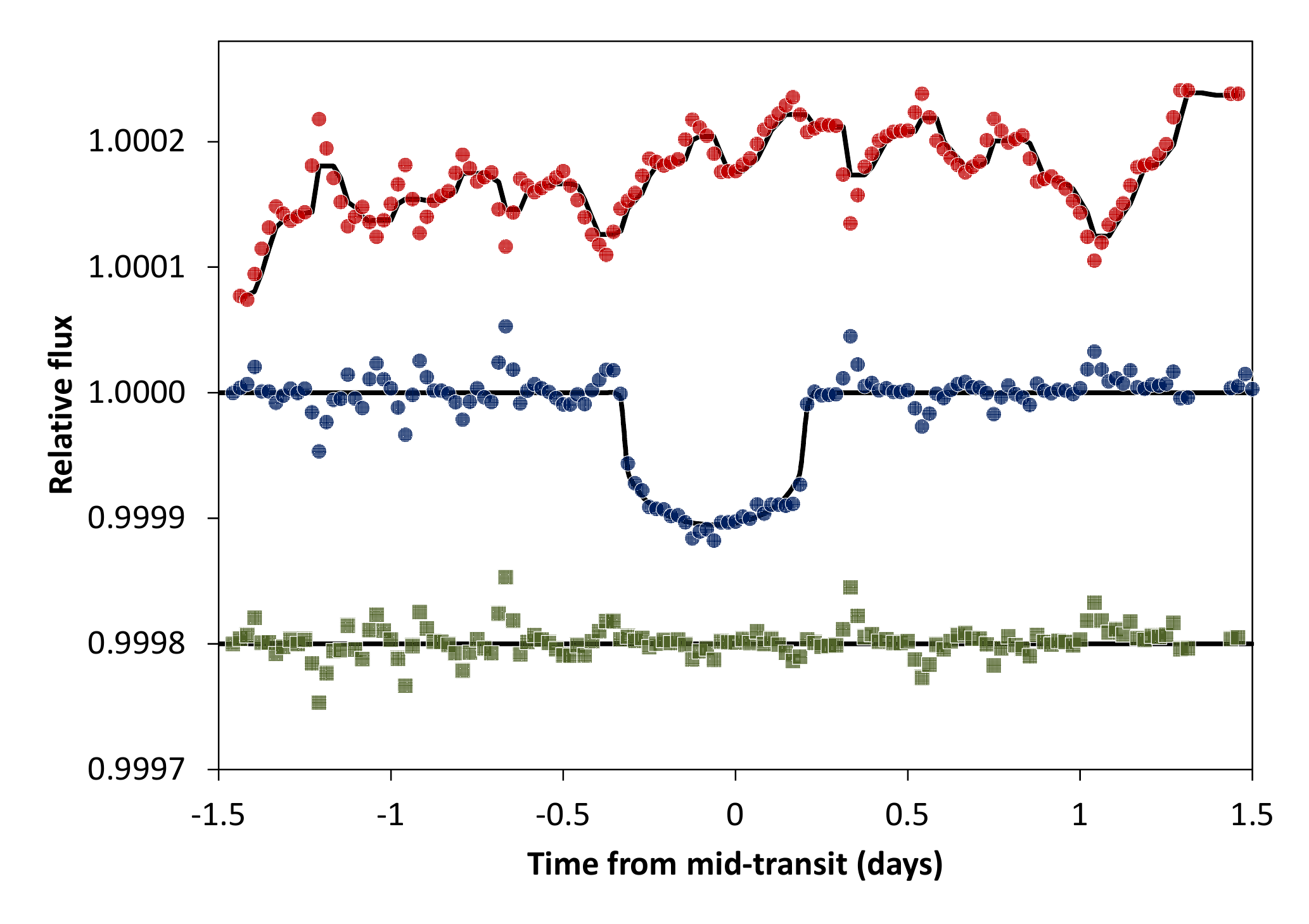}
\includegraphics[width=0.5\linewidth]{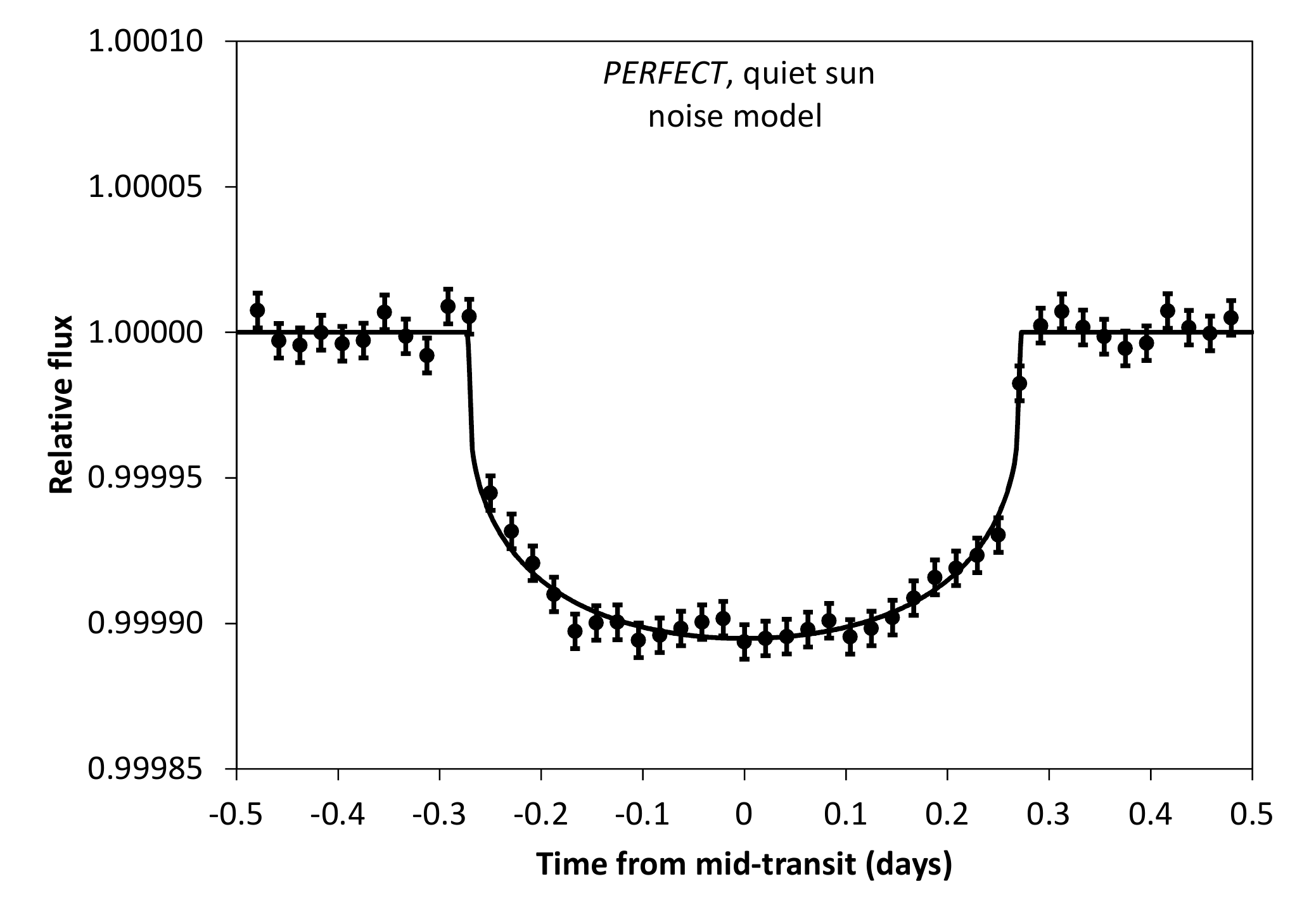}
\caption{\label{fig:earth}Earth observations after collecting data of six transits. Upper left: \textit{Kepler} observing the $K_{P}$=12 quiet sun (total noise: 26ppm). Middle left: The same observation for \textit{PLATO 2.0} (16ppm). Upper right: \textit{PLATO} 2.0 with the loud sun (29ppm). Middle right: \textit{PERFECT} observing the quiet sun (15ppm). The bottom panels show the result including our noise model, as explained in section~\ref{sub:noisemodelapplied}. The blue symbols on the bottom left, and the panel on the bottom right, are created using a \textit{simultaneous} fit of noise model and transit curve. The value for the quiet sun differs slightly from section~\ref{sub:instrumentalnoise} at this specific transit time. For better comparison, all axes are the same as in Figure~\ref{fig:mercury} for Mercury, Venus and Mars.}
\end{figure*}

\begin{figure}
\includegraphics[width=\linewidth]{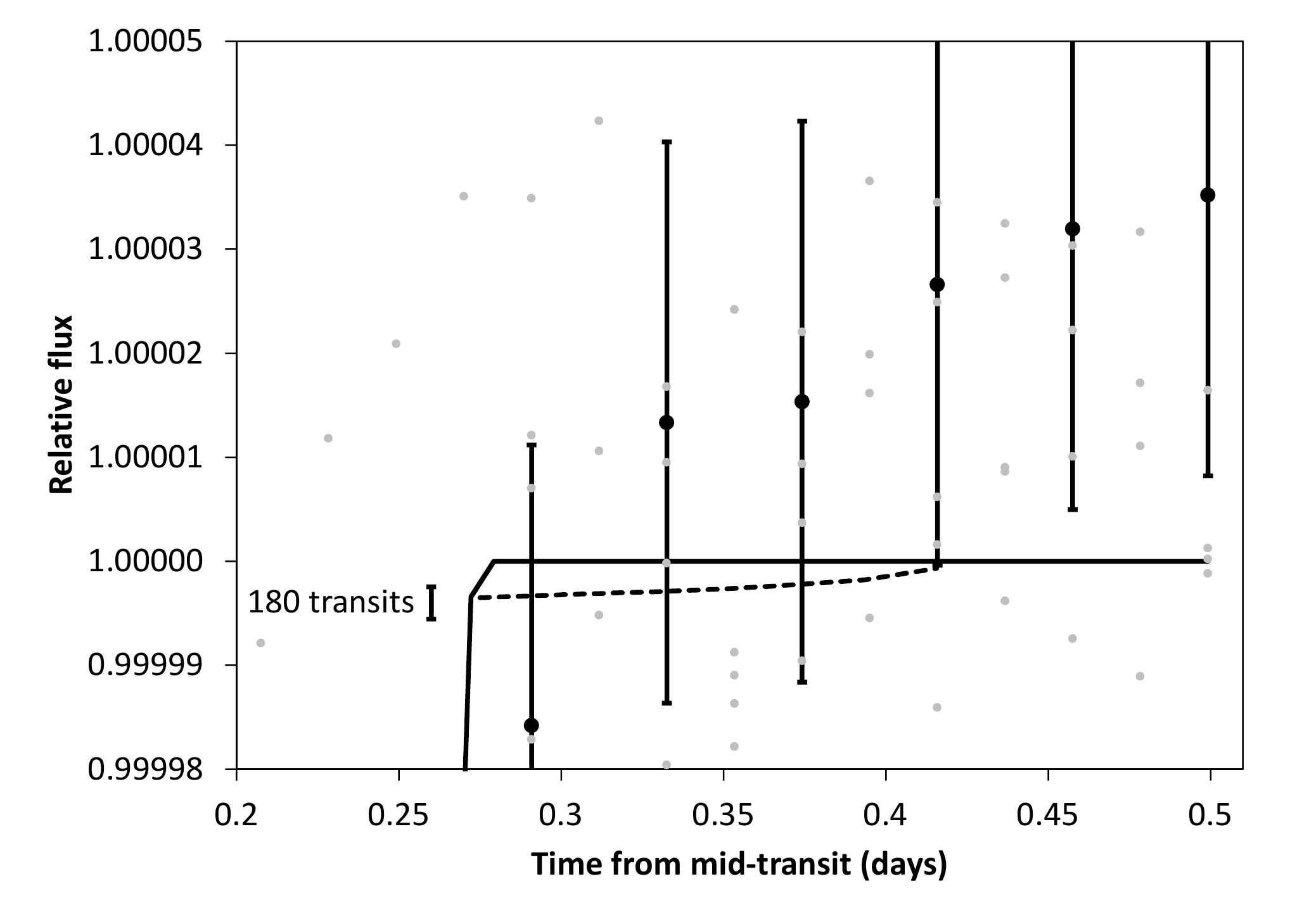}
\caption{\label{fig:luna}Orbital sampling effect (dashed line) for Earth's moon using \textit{PLATO 2.0} and the six quiet years of the solar cycle (2005--2010). Luna cannot be recovered due to (mainly stellar) noise. To achieve a 2$\sigma$ detection, $\sim$180 transits would be required.}
\end{figure}

\begin{figure*}
\includegraphics[width=0.5\linewidth]{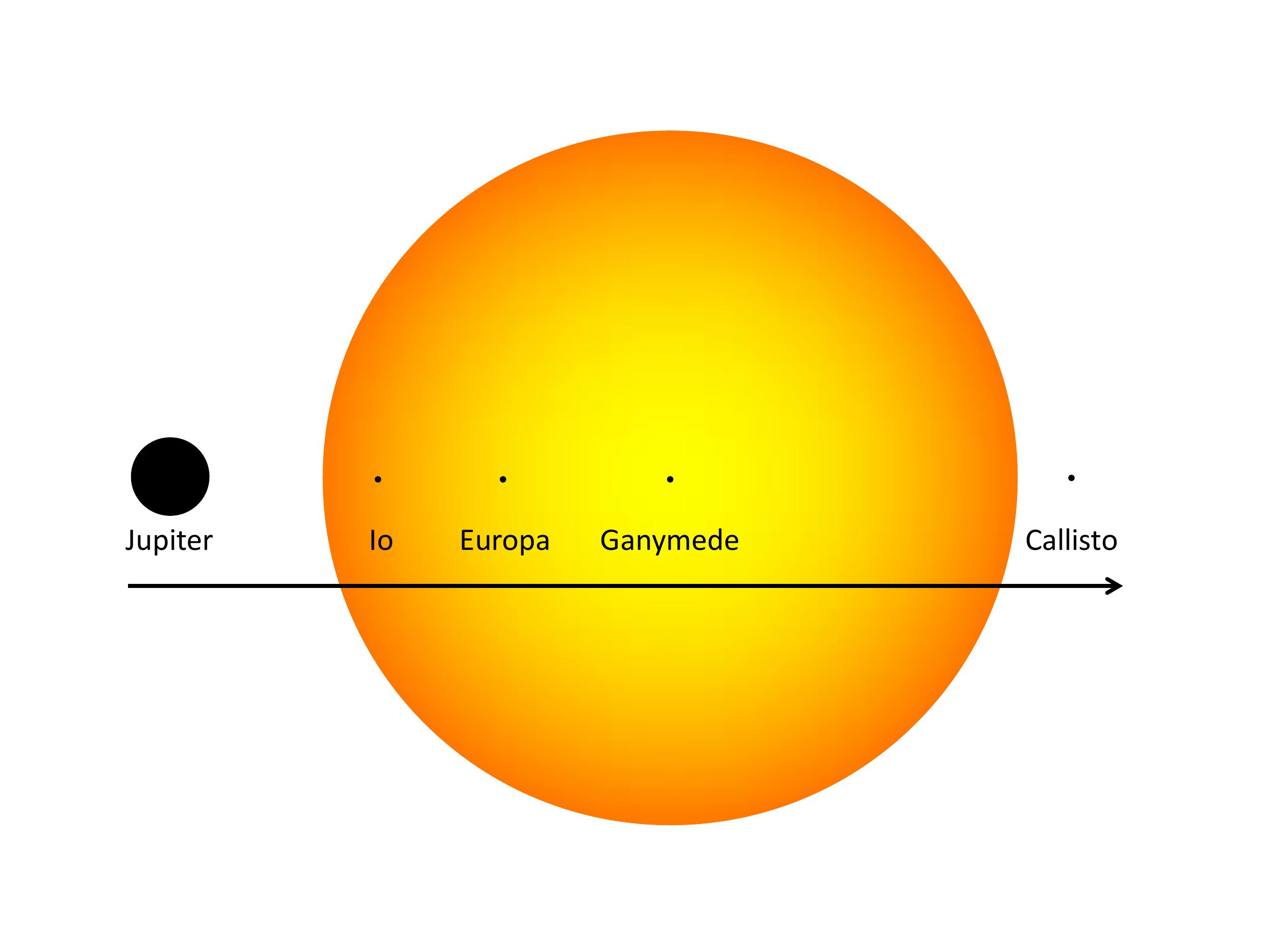}
\includegraphics[width=0.5\linewidth]{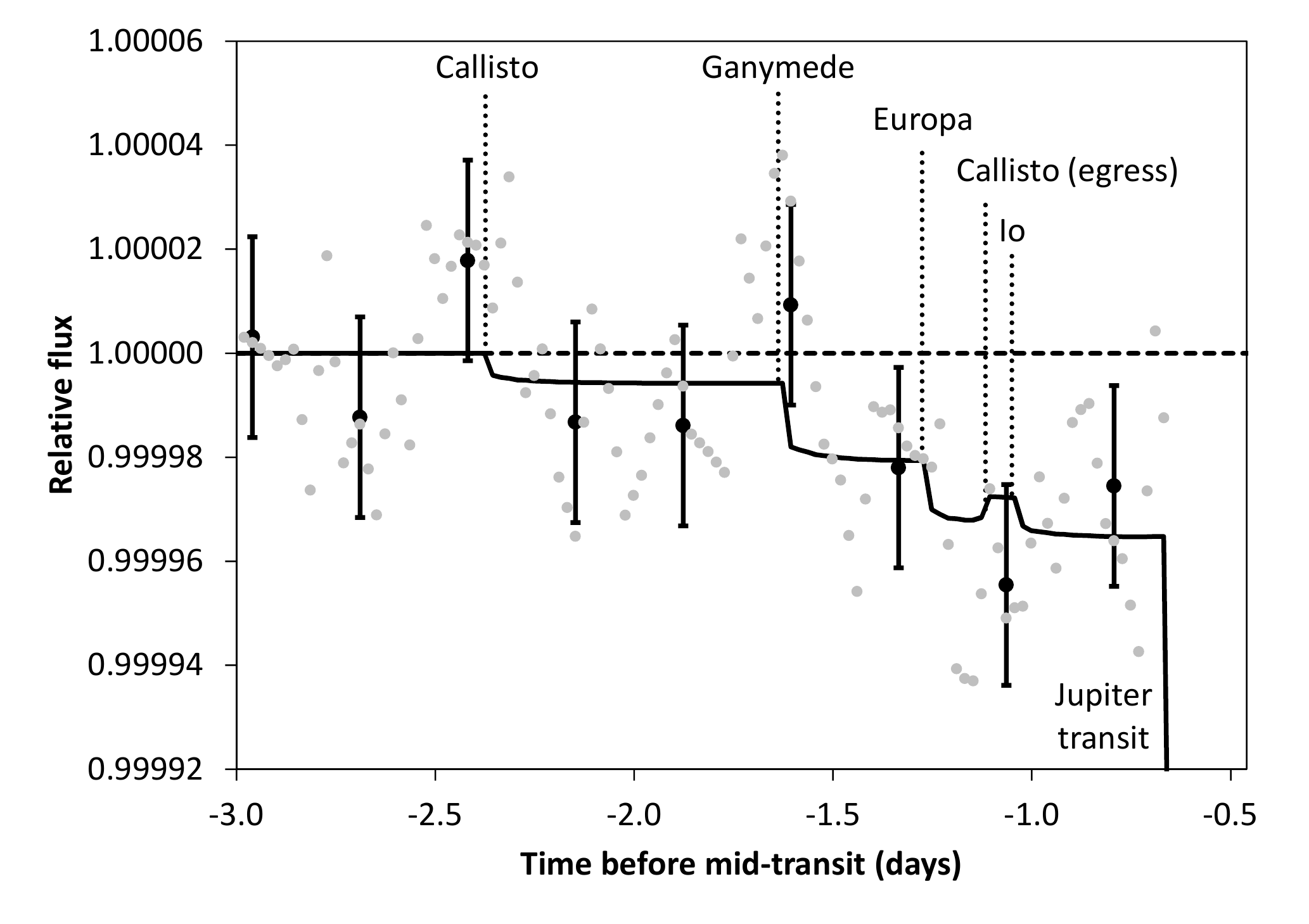}
\caption{\label{fig:jupisys}Left: Transit configuration of Jupiter system, with all moons at maximum separation on the ingress side. When Ganymede is at mid-transit, Jupiter has not started its transit yet, and Europa has already completed its own. Right: Flux for this configuration (line). Dots are real data LC bins (30min), data points with error bars binned in 6.5hrs. Note how the Callisto egress occurs before planetary ingress, and before Io's ingress. At no time, all four moons are transiting simultaneously, when at maximum separation.}
\end{figure*}

\subsection{Earth and moon}
\label{sub:earth}
As explained in section~\ref{sub:stellarnoise} and shown in Figure~\ref{fig:sunjitter}, red noise from stellar variation prevents the secure detection of an Earth-sized planet around a quiet G2-dwarf \textit{in a single observation}, no matter how good the photometer is. Also, single transit-like events can occur from sudden instrumental sensitivity drops, as described for the false-positive moon of Kepler-90g \citep{Kipping2015}. The situation improves when stacking a few transits. For an expected mission duration of six years for \textit{PLATO 2.0}, six transits could be observed of an Earth-analogue in the best case. As can be seen in Figure~\ref{fig:earth}, the detection is then possible at high confidence for both the active (S/N=15.5) and quiet (S/N=28.6) sun.

This finding is in agreement with the results of \citet{Aigrain2004}, who found that a single Earth transit cannot be detected, but a stack of six transits finds the planet at ``high'' S/N. While this result was achieved from synthetic data, \citet{Carpano2008} have also used \textit{VIRGO/DIARAD} data and concluded that planets smaller 2$R_{\oplus}$ cannot be detected in a single observation. This shows that our method is valid and consistent with previous findings.

The situation is equally promising for radial-velocity measurements (RV), assuming favorable conditions (single planet edge-on in circular orbit). As discussed by \citet{Lagrange2010} and \citet{Meunier2010}, an Earth analogue can be detected with 1\% false alarm probability after collecting $>200$ epochs spread over 4 years.

For the detection of Earth's moon, however, prospects are much worse. The tiny dip (6.6ppm) caused by Luna ($R_{\leftmoon}=0.27R_{\oplus}$) is invisible in a single transit. The transit timing variation (TTV) amplitude is 112 seconds, and the transit duration variation (TDV) only 14 seconds \citep{Kipping2009}. Errors from \textit{Kepler} are on the order of 400 to 800 seconds (S/N$\sim$0.3), making a detection of this configuration impossible. As TDVs, and their detection sensitivity, are a function of the impact parameter, there might be a few cases that are more fortunate. As explained by \citet{Kipping2009b}, for very high impact parameters $b>1$, i.e. grazing transits, TDVs are several times higher, up to 350s for realistic cases (see their Figure 2). Such candidates are hard to find, as the V-shaped transit curves are usually rejected by the automatic routines as eclipsing binaries. This possibility however, together with shorter integration times to improve timing sensitivity, might make the detection of exomoons, based on timings, possible with \textit{PLATO 2.0}. It seems crucial to have integration times as low as technically possible (ideally less than a minute), and improvements in algorithms are required to \textit{not} reject all V-shaped transits.

Without noise modeling, finding an Earth's moon analogue is also difficult using the \textit{orbital sampling effect} (OSE) as first described by \citet{Heller2014}. In short, when adding up many randomly sampled observations, a photometric flux loss appears in the phase-folded transit light curve, reflecting the moon's blocking of light. The effect depends mostly on the moon's radius and planetary distance. The OSE can be used to detect a significant flux loss \textit{before and after} the actual transit (if present), which might be indicative of an exomoon in transit. The basic idea is that at any given transit the moon(s) must be somewhere: They might transit before the planet, after the planet, or not at all -- depending on the orbit configuration. But by stacking many such transits, one gets, on average, a flux loss before and a flux loss after the exoplanet transit. While the OSE has shown to be useful when stacking many transits \citep{Hippke2015}, the sheer number of $\sim$180 transits required in this case (for a 2$\sigma$ detection) renders the method useless for Earth and moon (Figure~\ref{fig:luna}).

\subsection{Earth transit with noise modeling}
\label{sub:noisemodelapplied}
To explore the benefits of a tailor-made noise modeling, we have implemented a standard wavelet-based filter as presented by \citet{Carter2009} and explained in section~\ref{sub:noisemodel}. We simulate our Sun's noise as a stationary process plus a time-correlated process of spectral power density $1/f^\gamma$. To estimate the best global parameters, we used Monte-Carlo runs to minimize the post-model squared residals, but also restricted the parameters to not fit trends on times $<2$hrs in order to avoid over-fitting. The danger of such noise models is that with enough and sufficiently small parameters, one can fit out anything perfectly, be it a small transit (e.g. a moon), or a spot. For solar data, we find that the noise model is robust to filter out trends on timescales longer than a few hours, and more than a few 10s of ppm, so that an Earth-like transit signal benefits from noise modeling. Blindly recovering Earth's moon, however, doesn't work even with such a noise model, as the moon's transit signal (6.3ppm) is either still buried in noise, or has been removed altogether by too aggressive over-fitting. 

We show the application of our noise model in the bottom left panel of Figure~\ref{fig:earth}. The uppermost (red) symbols show the raw data without an injection, and the simulated (modeled) wavelet-filter as the black line behind the datapoints. After injection of the transit signal, we have then performed a \textit{simultaneous} fit to the transit curve and noise-model (blue symbols). The residuals (green) are shown in the bottom; these can be compared the the raw data (red). We have used the same solar data as in Figure~\ref{fig:sunjitter}, so that one can visually compare the benefit of the noise model.

In the bottom right panel of Figure~\ref{fig:earth}, then, we zoom into a single transit, offering $\sim2\times$ improved S/N. This shows the benefit of a noise model: It removes most of the time-correlated noise, and increases S/N at the same time. However, it is also suspect to removing smaller signals, such as moons. Also, it requires considerable effort to implement a model, validate its parameters, and apply it to the data. In the future, we must hope to achieve further theoretical advantages so that such models can be created automatically during planet-searches.

\subsection{Jupiter and moons}
\label{sub:jupiter}
A single Jupiter transit produces a deep (1.125\% = 11,250ppm), long (33hrs) transit dip, resulting in a highly significant (S/N=4,360) detection. Due to its long period of 11.86 years, we can not expect to observe a second transit with \textit{PLATO 2.0} during the spacecraft's expected 6-year lifetime. Assuming that the single transit itself is spotted in the data, we can search for accompanying moons. In the case of Jupiter, all of its larger moons are almost sky-coplanar and would thus also be transiting. In this example, we examine the case of all moons being at maximum separation, as this case is usually expected to yield the highest detection probability \citep{Hippke2015}. For our Jupiter, this is not the case: due to the large ($25R_{P}$) separation of Callisto, this moon finishes its transit (just) before Io goes into transit; there is no cumulative dip of all moons transiting simultaneously (when at maximum separation). We neglect all other small moons, as they contribute less than 1\% of additional transit depth.

As can be seen in Figure~\ref{fig:jupisys}, the whole transit ensemble can be detected with marginal confidence at S/N=8.6. Individual moons are not discernible, and it cannot be decided whether a single (large) moon, or a multiple moon configuration is observed. Such an observation would constitute a strong moon \textit{candidate} worth follow-up observations, but likely no clear detection.

\begin{figure*}
\includegraphics[width=0.5\linewidth]{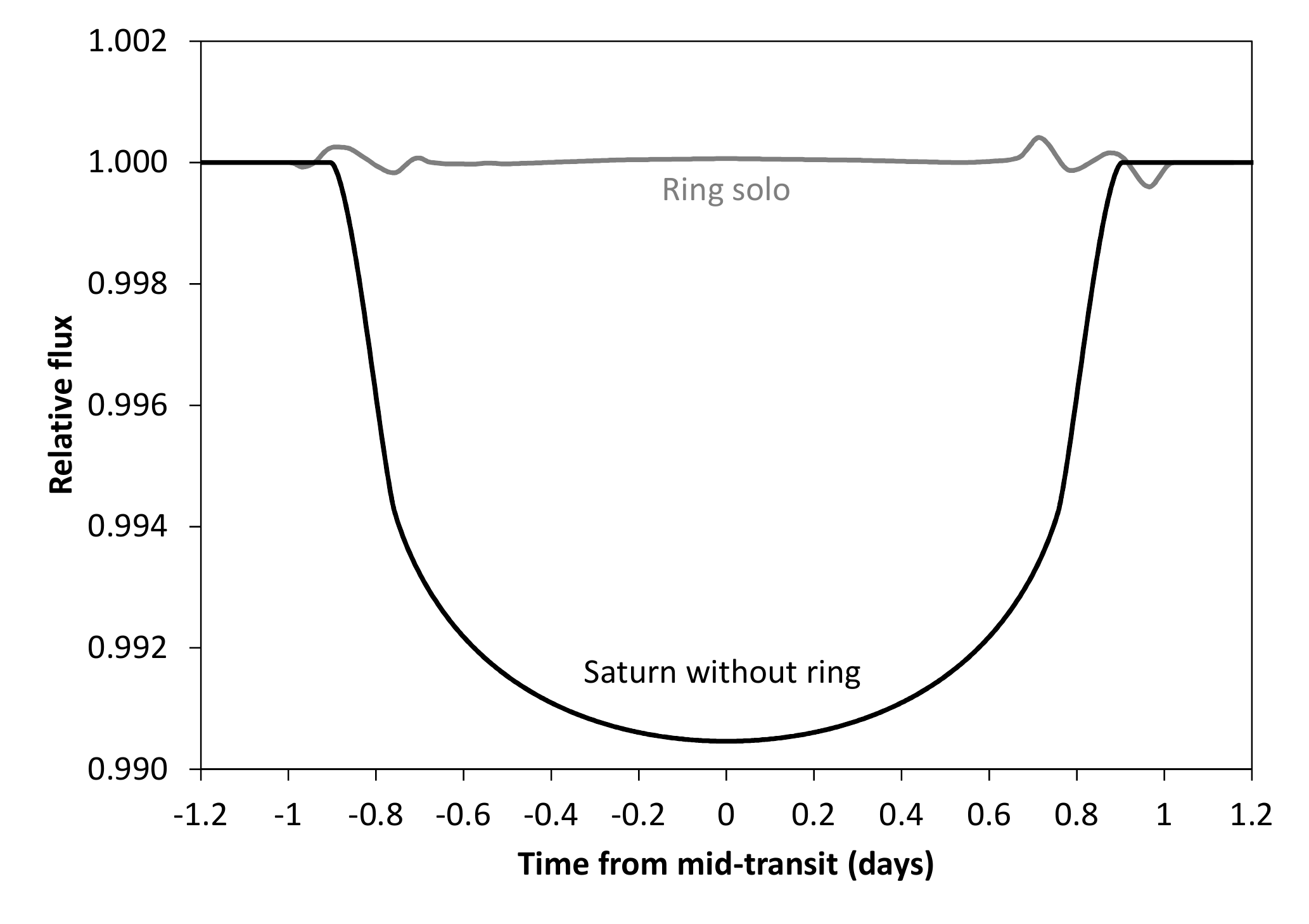}
\includegraphics[width=0.5\linewidth]{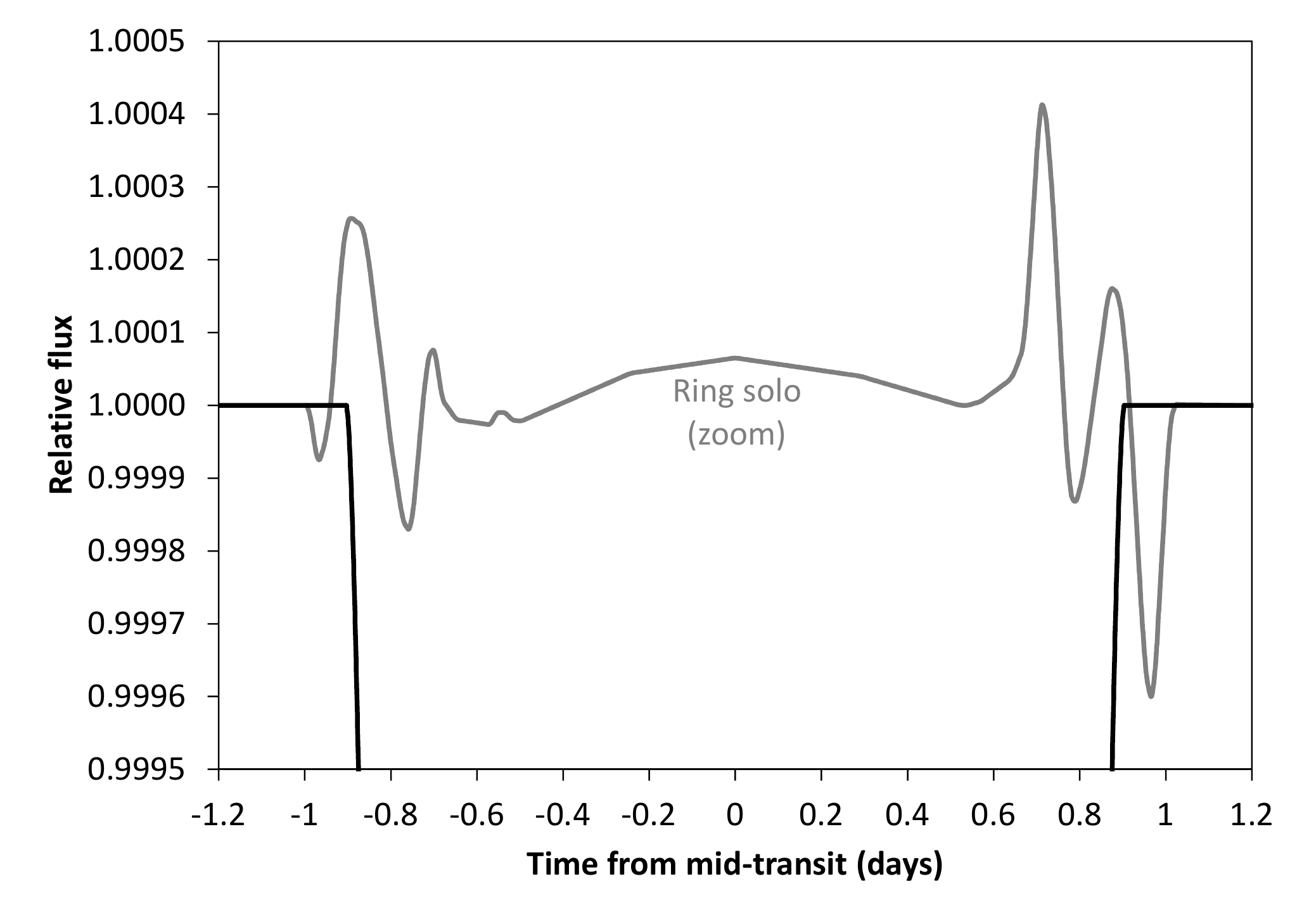}

\includegraphics[width=0.5\linewidth]{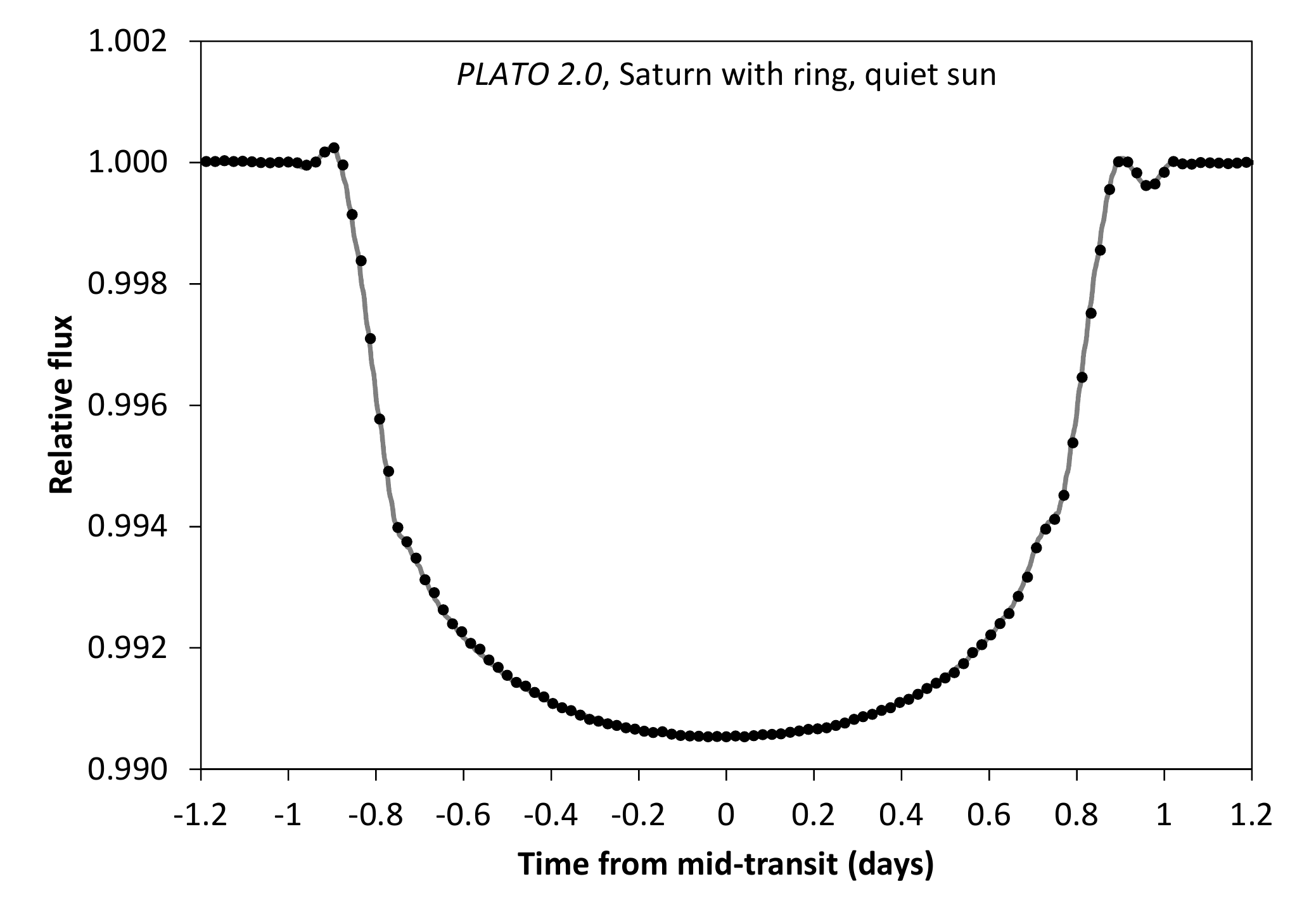}
\includegraphics[width=0.5\linewidth]{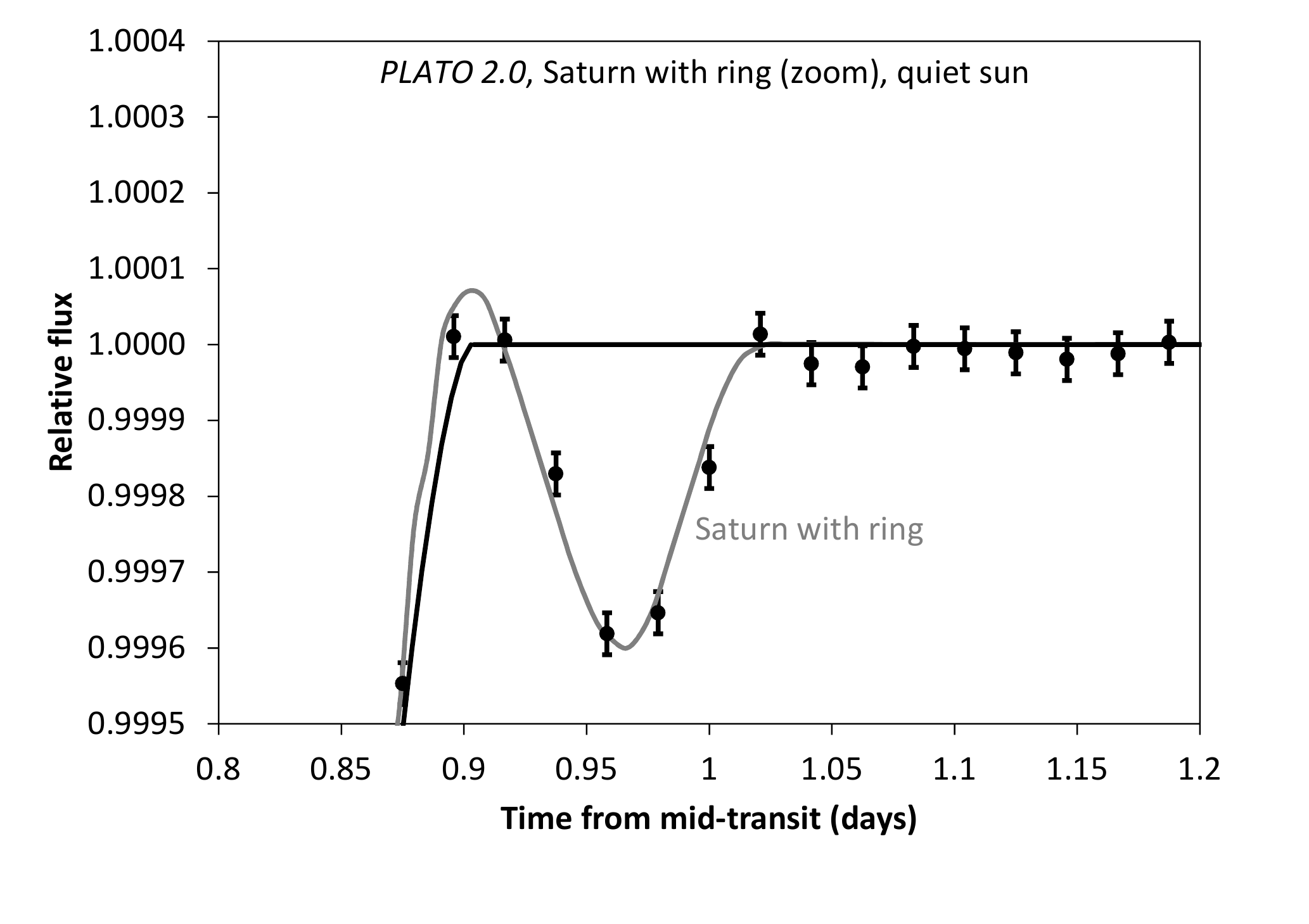}
\caption{\label{fig:saturnring}Upper left: Saturn transit without ring (black line) and theoretical ring solo transit (grey line). Upper right shows zoom. Higher than nominal flux is caused by diffractive forward-scattering of the rings. Lower left: Saturn including ring (black line) and data for \textit{PLATO 2.0} during quiet sun (errors as dot size). Lower right: Zoom into egress, where the ring is clearly detected. Data points are 30min LC bins.}
\end{figure*}

\subsection{Saturn with rings}
\label{sub:saturn}
In this section, we examine \textit{our} Saturn transiting \textit{our} quiet Sun observed through \textit{PLATO 2.0}. A simplified test for this configuration has been performed by \citet{Tusnski2013}, using synthetic \textit{Kepler} data and a dark Saturn-model. In the following, we use the real solar data and the model from \citet{Barnes2004}, which includes diffractive forward-scattering \citep{Dyudina2005}. For our Saturn, data are available from the 1989 occultation of 28 Sgr by Saturn \citep{French2000}, to adjust the model to the rings' complex nature (see Figure 8 in \citet{Barnes2004}). The main difference to the pure black transit is the flux \textit{gain} due to scattered light. We inject these data, together with the standard transit shape, and show the result in our Figure~\ref{fig:saturnring}. We neglect any moons, as they have been treated in the previous section. As can be seen in the graph, the rings are clearly recovered, with the flux gain prominently seen. As can be seen in the figure, a time resolution of at least 30min is required, to avoid smearing of the light curve features. The out-of-transit features alone account for a S/N=21.5, fully sufficient to claim a detection without modeling. Full modeling is however encouraged, as $\sim$50\% of the flux delta occurs during transit, and most of it during ingress/egress. As pointed out by \citet{Zuluaga2015}, rings cause an increase in transit depth that ``may lead to misclassification of ringed planetary candidates as false-positives and/or the underestimation of planetary density''. By comparing results from astrodensity profiling \citep{Kipping2014} to those from astroseismology \citep{Huber2013}, such anomaly low density planets could be detected.

Saturn itself produces a highly significant S/N=3,575 in a single transit.

\begin{figure}
\includegraphics[width=\linewidth]{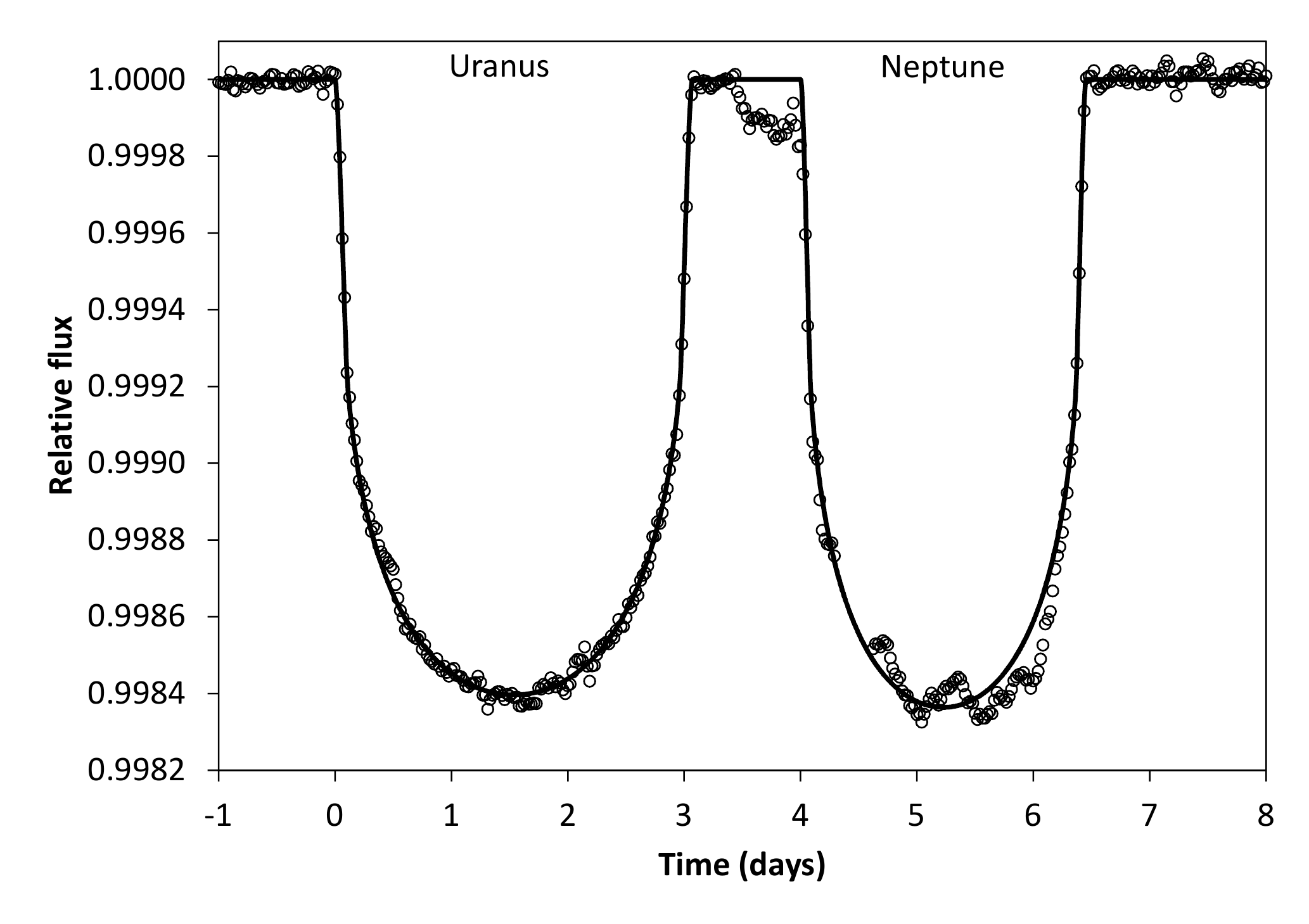}
\caption{\label{fig:uranusneptune}Uranus and Neptune transits with data as 30min LC bins. Neptune's transit is slightly shallower but longer. Just before Neptune's transit, a spot-induced 105ppm flux-drop occurs, which could be mistaken for a 1.1$R_{\oplus}$ exomoon.}
\end{figure}

\subsection{Uranus and Neptune}
\label{sub:uranus}
Uranus and Neptune produce deep (1,382ppm, 1,354ppm), long (2.45d, 3.07d) transits. The S/N for Neptune (783 vs. 714) is slightly higher, due to its longer transit duration. As can be seen in Figure~\ref{fig:uranusneptune}, the transit is visually compelling, but stellar noise may mimic moons where there are none. For a test, we have injected Neptune's largest moon, Triton (0.21$R_{\oplus}$, 4ppm), which is unrecoverable in the noise. Furthermore, Triton is usually not seen in transit, due to its inclination of 129.6$^\circ$ \citep{Agnor2006} with respect to the Laplacian plane of the solar system. In such cases, even the largest moons would remain unnoticed.

\begin{figure}
\includegraphics[width=\linewidth]{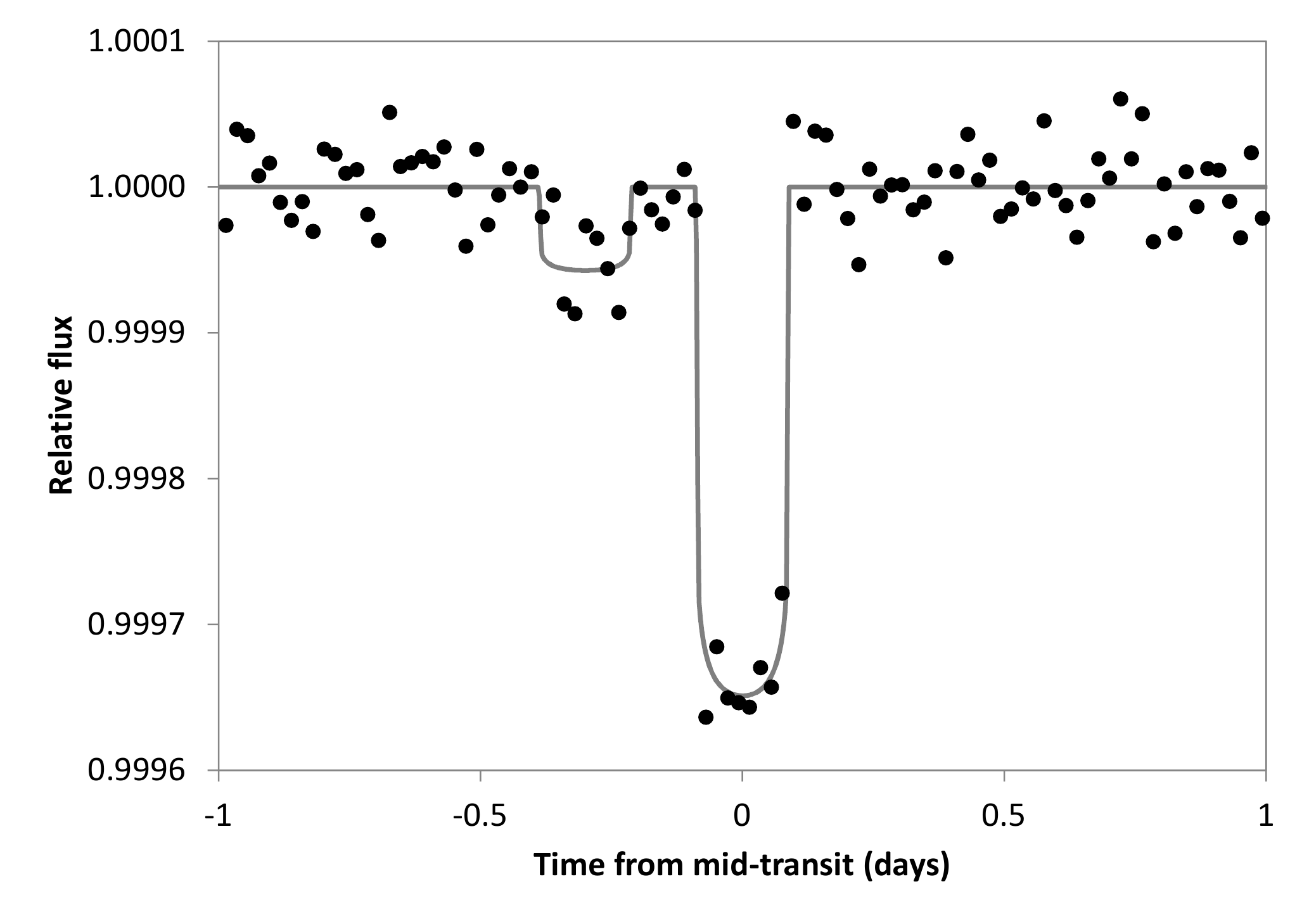}

\includegraphics[width=\linewidth]{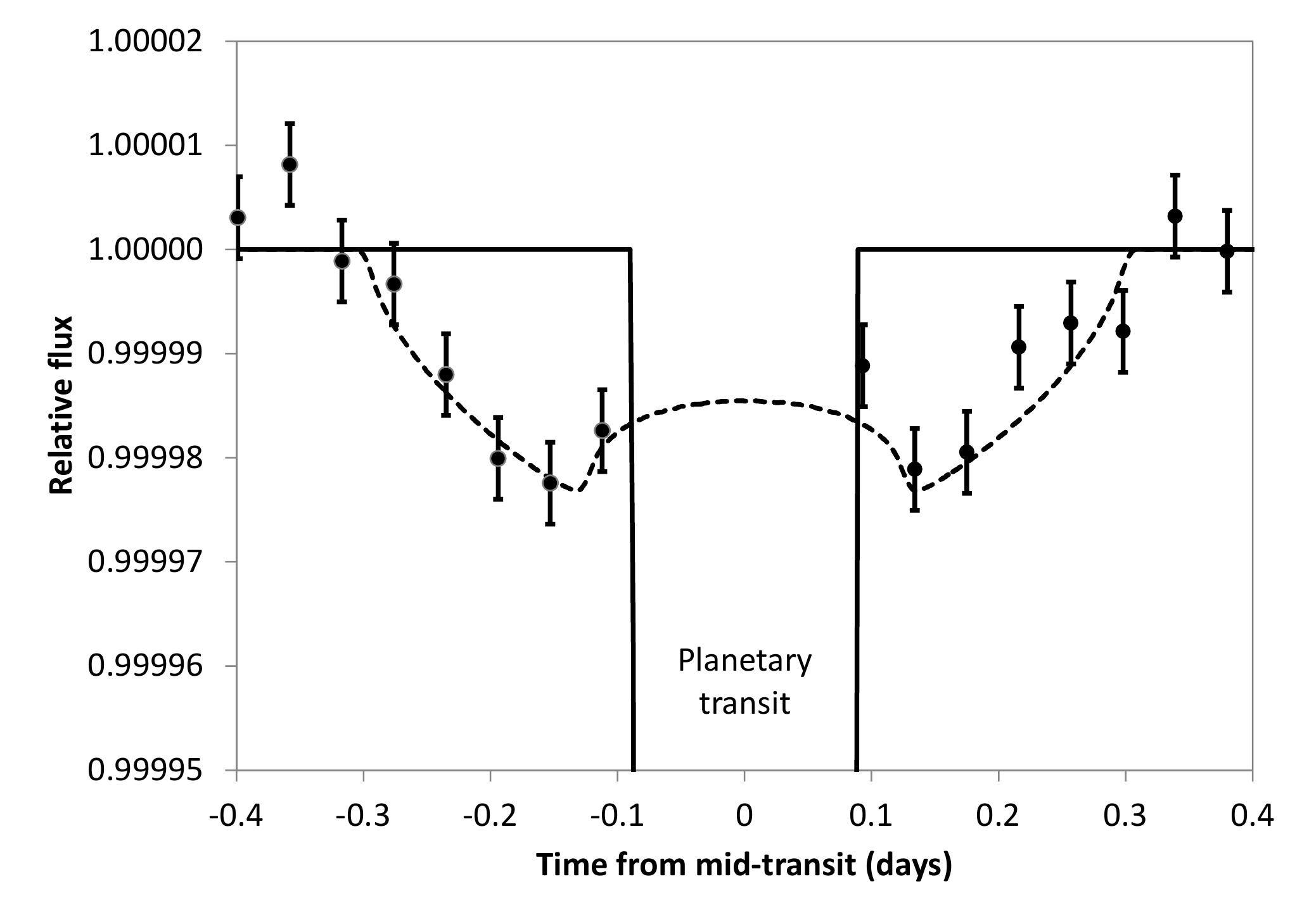}

\includegraphics[width=\linewidth]{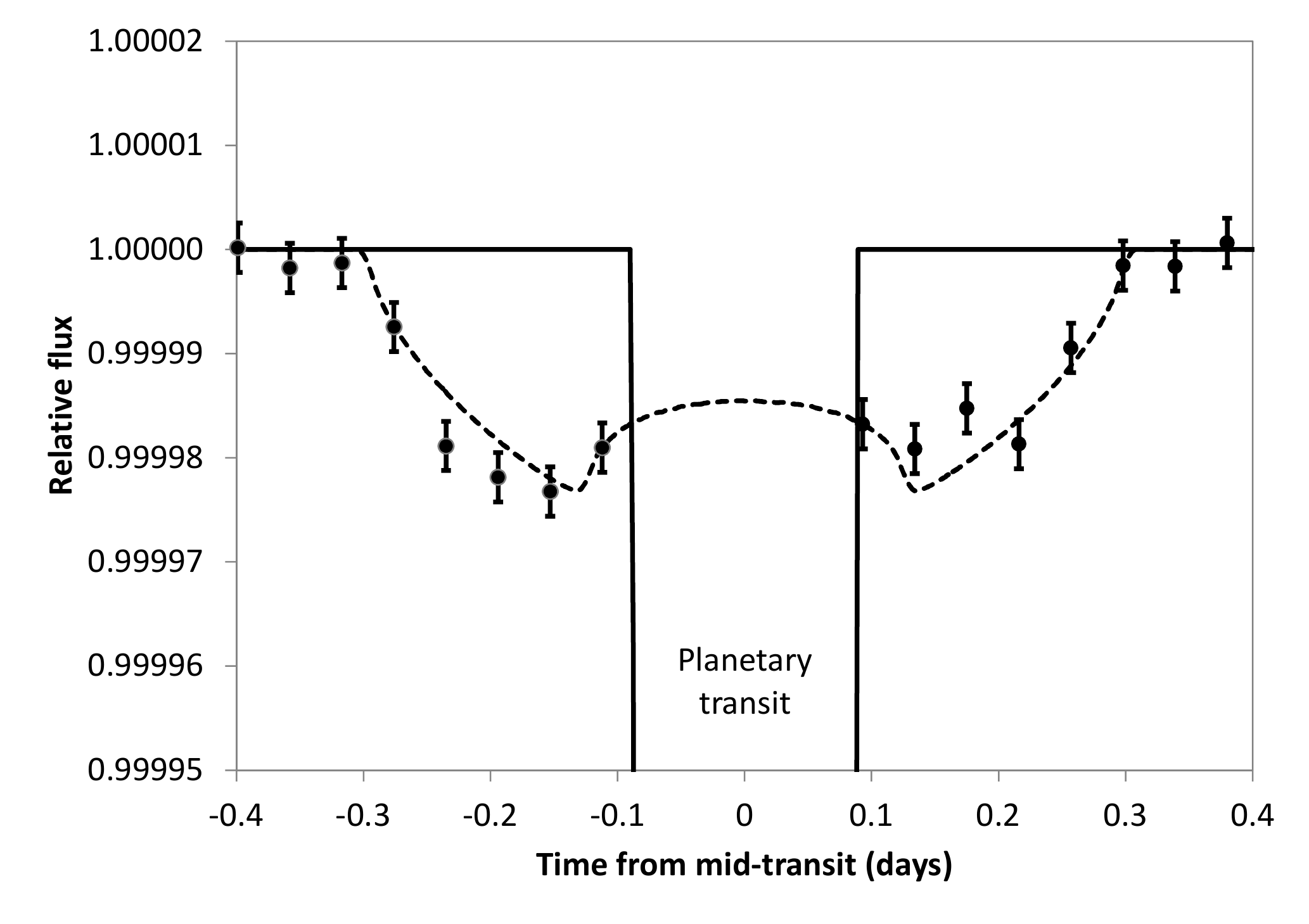}
\caption{\label{fig:m-dwarf}2.0$R\oplus$ planet and Ganymede-sized (0.4$R_{\oplus}$) moon orbiting a 0.5$R_{\odot}$ M-dwarf with P=23.97d, resembling Kepler-236c, but with low noise properties (stellar noise CDPP=7ppm). Top: Single transit with \textit{PERFECT} giving a clear dip for the planet (S/N=22.8), but low S/N=3.9 for the moon at maximum separation. Middle: After stacking 6yrs (100 orbits) with \textit{PLATO 2.0}, the orbital sampling effect clearly recovers the moon. The gain from \textit{PERFECT} is in this case $\sim$50\% (bottom).}
\end{figure}

\begin{figure}
\includegraphics[width=1.1\linewidth]{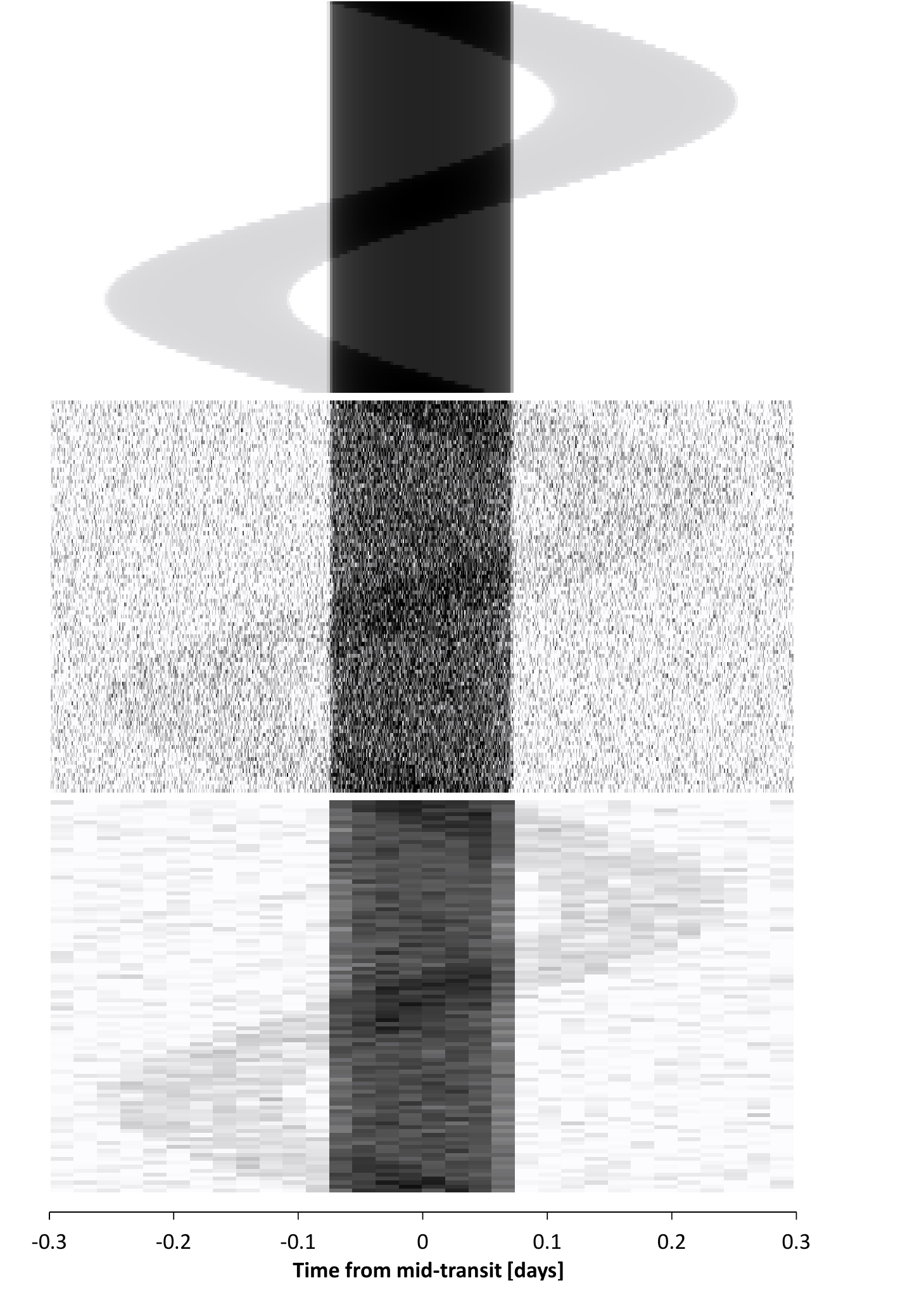}
\caption{\label{fig:river}River plot of moon injection, compare Figure~\ref{fig:m-dwarf}, using the \textit{PERFECT} data. Dark colors are lower flux levels and indicate the transits. The horizontal axis of each plot represents one transit event, with the vertical dark stripe being caused by the planetary transit. Each of the 100 rows is one period. Top: Calculated injections. Middle: One-minute sampling. Bottom: 30-minute sampling.}
\end{figure}

\subsection{Super-Earth and moon transiting the M-dwarf}
\label{sub:mdwarf}
We have now completed the tour through our solar system, and have seen that detecting planets and rings is possible for many scenarios. Exomoons, however, are more difficult to observe, making a detection unlikely for solar system analogues, as explained in sections \ref{sub:earth} and \ref{sub:jupiter}. We will thus try a more promising configuration: A very quiet M-dwarf (0.5$R_{\odot}$, stellar noise: 7ppm CDDP), assumed to have a transiting planet of Super-Earth size (2.0$R_{\oplus}$) in a P=23.97d orbit, with an accompanying Ganymede-sized (0.4$R_{\oplus}$) moon. This system is inspired by Kepler-236c and shares its stellar and planetary size, and their separation. The stellar noise properties are as described in section \ref{sub:target}, resembling the M-dwarf KIC7842386 with 6.7ppm Gaussian noise on 6.5hrs timescale. As can be seen in Figure~\ref{fig:m-dwarf}, a single transit is clearly detected for the planet, but not for the moon. With \textit{PLATO 2.0} photometry and 6 years of data (100 orbits), the moon dip is clearly retrieved through its orbital sampling effect (middle plot). The gain from \textit{PERFECT} (bottom plot) is in this case $\sim$50\%, thanks to the high share of instrumental noise (7ppm stellar, 10ppm \textit{PLATO 2.0}). Interestingly, \textit{PLATO 2.0} photometry allows for the detection of the ``cap'' in the OSE. This feature represents the relative flux gain before and after planetary transit, when a large-orbit moon goes into stacked ingress (egress) before (after) the planet transits. Detecting this feature gives more of a ``shape'' to the OSE, in contrast to a pure instrumental decrease in luminosity. To give an impression of the unfolded photometry, we show a riverplot \citep{Carter2012, Nesvorny2013} in Figure~\ref{fig:river}. With 100 transits (each row is one period) of \textit{PERFECT} data, the moon transit signature is visually evident even without folding. With \textit{PLATO 2.0} data, it is also visible when stretching the gray- scale. Such plots can be used for a sanity check of moons with \textit{PLATO 2.0} data.

\subsection{Oblate planets}
\label{sub:oblate}

\begin{figure}
\includegraphics[width=\linewidth]{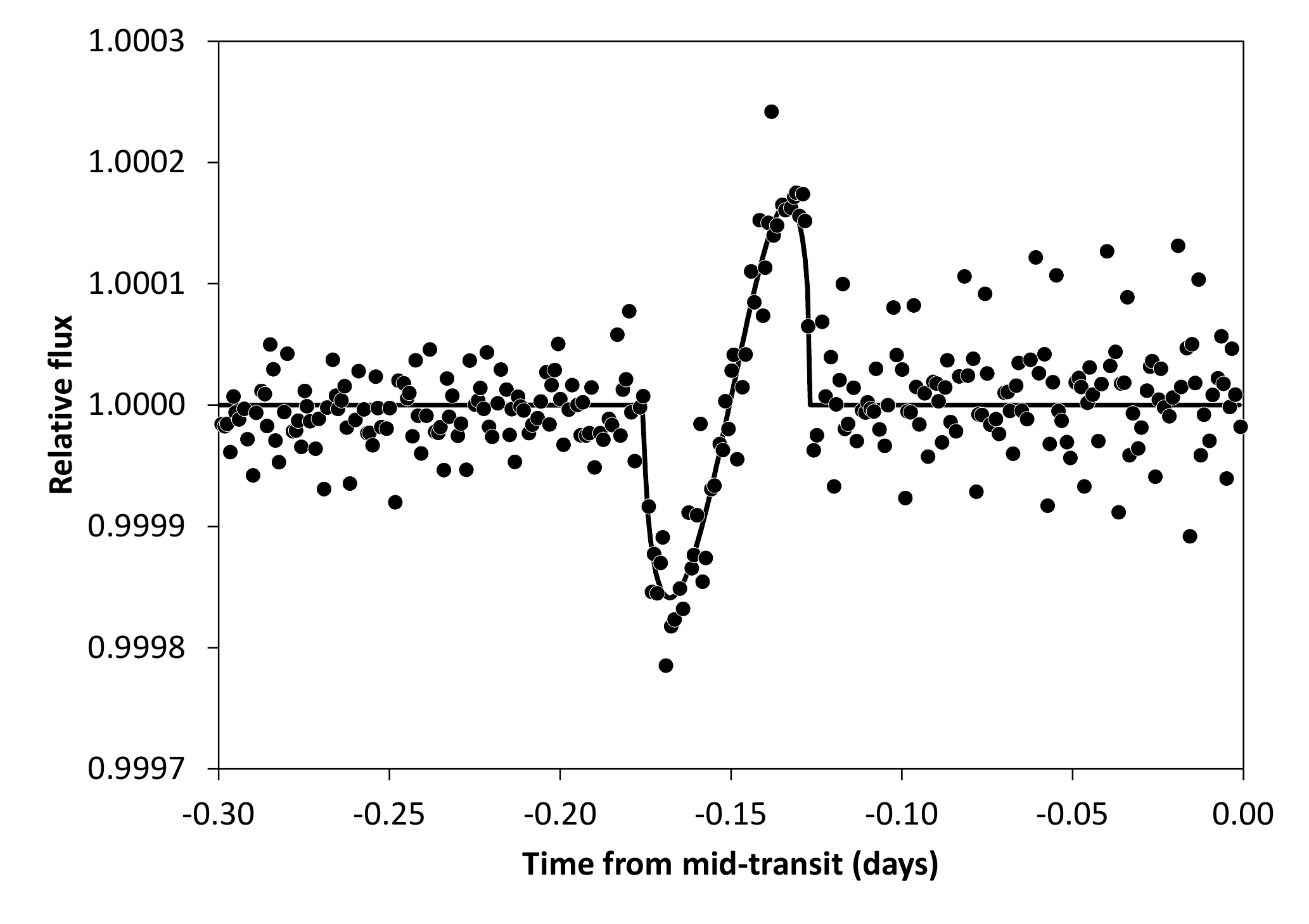}
\caption{\label{fig:oblate}Delta (line) between a spherical and an oblate Jupiter-sized planet, using Saturn-like ($f=0.1$) oblateness, spin obliquity of 45$^{\circ}$ and a 87-day (Mercury) orbit. Data points are for 6 years of quiet solar observations using \textit{PLATO 2.0}. No noise modeling.}
\end{figure}

Planets are never perfect spheres, but show oblateness to varying degrees. Measuring the oblateness of exoplanets will help our understanding of planetary formation, rotation, and internal structure \citep{Carter2010}. The transit light curve of an oblate planet is different from that of a spherical, with the main delta during ingress and egress. \citet{Zhu2014} have performed a search in \textit{Kepler} data and find the hot-Jupiter \mbox{HAT-P-7b} to be a good candidate. Their analysis shows that \textit{Kepler}-level photometry is sufficient to detect ``Saturn-like oblateness ($f=0.1$) for giant planets ($R_{P}/R_{*}=0.1$) around relatively bright (12 mag) stars''.

To analyze the prospects of \textit{PLATO 2.0} in this regard, we have injected a Jupiter-sized planet including Saturn-like ($f=0.1$) oblateness with a 87-day (Mercury) period into 6yrs of quiet solar data. The calculated oblateness delta is the one from \citet{Zhu2014} and has kindly been provided by the author (Wei Zhu 2015, priv. comm.). We chose a configuration that generates a large oblateness signal, using a projected planet spin obliquity of 45$^{\circ}$. As can be seen in Figure~\ref{fig:oblate}, the signal is clearly retrieved, at S/N=16. As for planetary transits, the signal-to-noise ratio will be higher for shorter periods, larger oblateness, and larger planet-to-star ratios. On the other hand, it will be harder to detect planets with smaller oblateness, longer periods, and smaller radii (for a constant projected planet spin obliquity). Using the large sample expected from \textit{PLATO 2.0}, we can hope to measure the oblateness of hundreds of planets, and create an insightful distribution statistic.

\section{Discussion}
After examining these transits, we should summarize the learnings. It is clear that future photometry will allow for many more discoveries, and we should prepare to get most out of these data.

\subsection{Observation strategies for \textit{PLATO 2.0}: Single transits}
\label{strategies}
While transit detections of medium-sized, short-orbit planets such as Earth and Venus will be feasible with future photometry, the stellar noise will put limits to planet sizes and configurations for Mercury and Mars analogues. The greatest challenge will be longer-orbit transits (cold gas giant analogues), for which only one transit can be observed in a reasonable ($<$10 years) time span. With their long periods of 84 years (Uranus) and 165 years (Neptune), the observation of a single transit for any given solar system analogue is unlikely ($P_{tr}=0.024\%$ for Uranus, $P_{tr}=0.014\%$ for Neptune). The occurrence rate of such planets is currently not well constrained, as only a few examples have been found through gravitational microlensing \citep{Furusawa2013,Sumi2010}, but is believed to be $\ge$16\% at 90\% confidence \citep{Gould2006}. \textit{PLATO 2.0} will observe 85,000 bright (V$<$11) stars \citep{Rauer2014}, and all of these can be expected to yield high (S/N$>500$) detection potential for cold Uranus and Neptune analogues. For the full sample of $>$1bn stars, one might expect at least half of them to give sufficient S/N ($>20$) for a clear detection of such planets. Consequently, it can be expected to detect a considerable number of single-transits events for such planets with \textit{PLATO 2.0}: With an occurrence rate of 16\%, and 0.024\% transit probability, and observing 6 out of 84 orbital years, we can expect to find one Uranus analogue among 365,217 stars. For a 500mn sample of \textit{PLATO 2.0} data, we can thus expect to find 1,369 Uranuses (and not a single one among \textit{Kepler's} 150,000 stars). 

The frequency of Jupiter analogues at distances of \mbox{3-6 AU} has been estimated as $5\pm2$\% by \citet{Lineweaver2003}, or $3.3\pm1.4$\% \citep{Wittenmeyer2011}. Assuming a 5\% frequency, we can expect to find $\sim$5,000 Jupiters and $\sim4,000$ Saturns, accounting for their higher transit probability (compared to Neptunes or Uranuses).

It is clear that these numbers are only very rough estimates, and might be subject to change by an order of magnitude. Through the very high number of stars observed with \textit{PLATO 2.0}, however, it will be possible to observe single-transit events of all solar system analogues. Current exoplanet science relies on multiple transits to determine periods, and confirm transit signals. It will be the challenge of these large-sample future missions to treat single events in a way that allow for their detection and confirmation. This will put our own solar system in perspective: Is a configuration of rocky inner, and gaseous outer planets common, or exotic?

\subsection{Validation of transit signals}
\label{sub:disc}
There are three principle ways to validate transit-like signals. The first is to check if they are, at all, astrophysically possible. If this is the case, we can try to eliminate those that are more likely to be caused by instrumental or stellar artifacts. 

The first check should be whether a putative star-planet-moon (or ring) configuration is astrophysically possible, plausible and stable over the long time. While the possibility of any configuration is essential, the situation is less clear for the plausibility: Before the discovery of Hot Jupiters, such planets would have seemed implausible, and their discovery \citep{Mayor1995} was challenged as they were found to be incompatible with theories of planetary formation \citep{Rasio1996}. 

This is also true for the stability. For example, we should assume that most moons are stable over long (Gyr) times, but there might be configurations for which this is not the case. A related example are Saturn's rings, which are known to be unstable on timescales of $<$100 Myr \citep{Dougherty2009}. Thus, instability should raise doubts about the presence of a moon (and less so about the presence of a ring), but does not proof their non-existence.

To check the plausibility of moons, we can require that prograde moons are within $\sim$39\% of their Hill radius, and retrograde moons to be within 93\% to be stable \citep{Domingos2006}. Also, they cannot have their orbit inside the Roche lobe, otherwise they would be torn apart and create a ring system. Similar arguments can be made for ring systems: When a flux increase (through diffractive forward-scattering) favours a ring composition of ice over rock, then such a ring planet cannot be too close to its host star, otherwise the ice would evaporate. Furthermore, we can count the fraction of transits that exhibit \textit{no} moon-like signals (assuming a coplanar moon), and compare this to the theoretical upper limit of $\sim$6.4\%, as derived by \citet{Heller2012} using the Roche stability criterion.

Sometimes, however, the situation is less clear. Recently, a single moon-like transit signal was reported by \citet{Cabrera2014} in the photometry of Kepler-90g. The best-fit parameters were a physically possible $7.96\pm0.65R_{\oplus}$ planet orbited by a $1.88\pm0.21R_{\oplus}$ exomoon. As reported by \citet{Kipping2015}, the signal is well explained by an instrumental sensitivity drop of a single pixel on the CCD. The authors introduce a method dubbed ``centroid map'', which compares the sensitivity of neighboring CCD pixels over time. This method allows to attribute probabilities to instrumental changes in sensitivities, for example caused by cosmic ray hits.

While such sensitivity drops are severe, but usually short events, there might also be instrumental trends on timescales longer than the actual transit(s). It has proven effective to remove these by cross-correlating them with their physical originator (when the data is available), for example temperature drifts, or shifts on the CCD have been corrected with proxy data for Hubble Space Telescope data \citep{Demory2015}, and in the \textit{Kepler} K2 mission \citep{Vanderburg2015,Foreman2015}.

When it comes to stellar noise, such proxy data is usually not available, but could be obtained from observations in multiple wavelengths through filters, or spectra. Stellar rotation can be measured  \citep[e.g.][]{McQuillan2013} and used to apply a suitable noise model. A few cases have been reported where transit depth variations might be explained by the stellar rotation phase which causes the transiting planet to occult star spots on a highly spotted star \citep{Croll2015}. With a combined detection of TTVs and rotation, it is also possible to distinguish between prograde and retrograde motion \citep{Holczer2015}.

These measurements make actual use of the otherwise unwanted stellar variation. In order to only distinguish between stellar flux variation and a transit signal, statistical tools can be used. We suggest to compare the occurrence rate of the signal(s) in question, including their shape, to the whole dataset in suitably chosen time bins. Ideally, there should be no other features in the whole dataset compared to the one in question. To validate exomoon signals, \citet{Hippke2015} argued that a test should be made, checking whether it is possible to shift the (virtual) folded transit time (to any other position), and then still have a significant dip. A variation of this test can check the uniqueness of any given signal.

\subsection{Data handling burdens}
Given the required detection techniques, it will be another challenge to efficiently mine the large data volume expected by \textit{PLATO 2.0}. When extrapolating the \textit{Kepler} data volume ($\sim$16 MByte per star and year in 1min integrations), we will have to search 48 PByte ($48\times10^{15}$ Byte) of data. The storage of these data alone costs $\sim$480,000 USD, in 2015 storage prices. Hard disk capacities have increased by a factor of 16 (for constant nominal prices)\footnote{\url{http://www.jcmit.com/diskprice.htm}, retrieved 09-Mar 2015} between 2004 and 2014. Extrapolating to the \textit{PLATO 2.0} mission in 2024 will estimate storage requirements of $\sim$30,000 USD. Clearly, most analyses will have to be done remotely, with the data stored in central facilities such as the Mikulski Archive for Space Telescopes (MAST)\footnote{\url{https://archive.stsci.edu/}}. Searches in \textit{Kepler} data today can be done on the researcher's own machine, as the total volume of the dataset is $<1$TB, i.e. less than 100 USD in 2015 prices. Depending on the computational demands of a specific research, processing is done on single computers, University Clusters, or the NASA Supercomputer. We expect that most or all large-scale searches in \textit{PLATO 2.0} data will need to be performed remotely, using super-computer facilities.

\begin{table}
\small
\caption{S/N for solar system transit objects, using \textit{PLATO 2.0} and 6yrs of quiet sun data\label{tab:sntable}}
\begin{tabular}{lccl}
\tableline
Object                                           & \# transits         & S/N  & Comment \\ 
\tableline
$\mercury$  Mercury                              & 25 & 7.0     &  \\
$\venus$    Venus                                & 9  & 23.8    &  \\
$\oplus$    Earth                                & 6  & 28.6    &  \\
\noindent\hspace*{1mm} $\leftmoon$ Luna          & 6  & 2.1     &  \\
$\mars$ Mars                                     & 3  & 7.2     &  \\
$\jupiter$ Jupiter                               & 1  & 4,360   &  \\
\noindent\hspace*{1mm}  $^{....}$ Galilean moons & 1  & 8.6     & out-of-transit \\
$\saturn$ Saturn                                 & 1  & 3,575   &  \\
\noindent\hspace*{1mm} $\ominus$ Saturn's rings  & 1  & 21.5    & out-of-transit \\
$\uranus$ Uranus                                 & 1  & 714     &  \\
$\neptune$ Neptune                               & 1  & 783     &  \\
\tableline
\end{tabular}
\end{table}

\begin{figure}
\includegraphics[width=\linewidth]{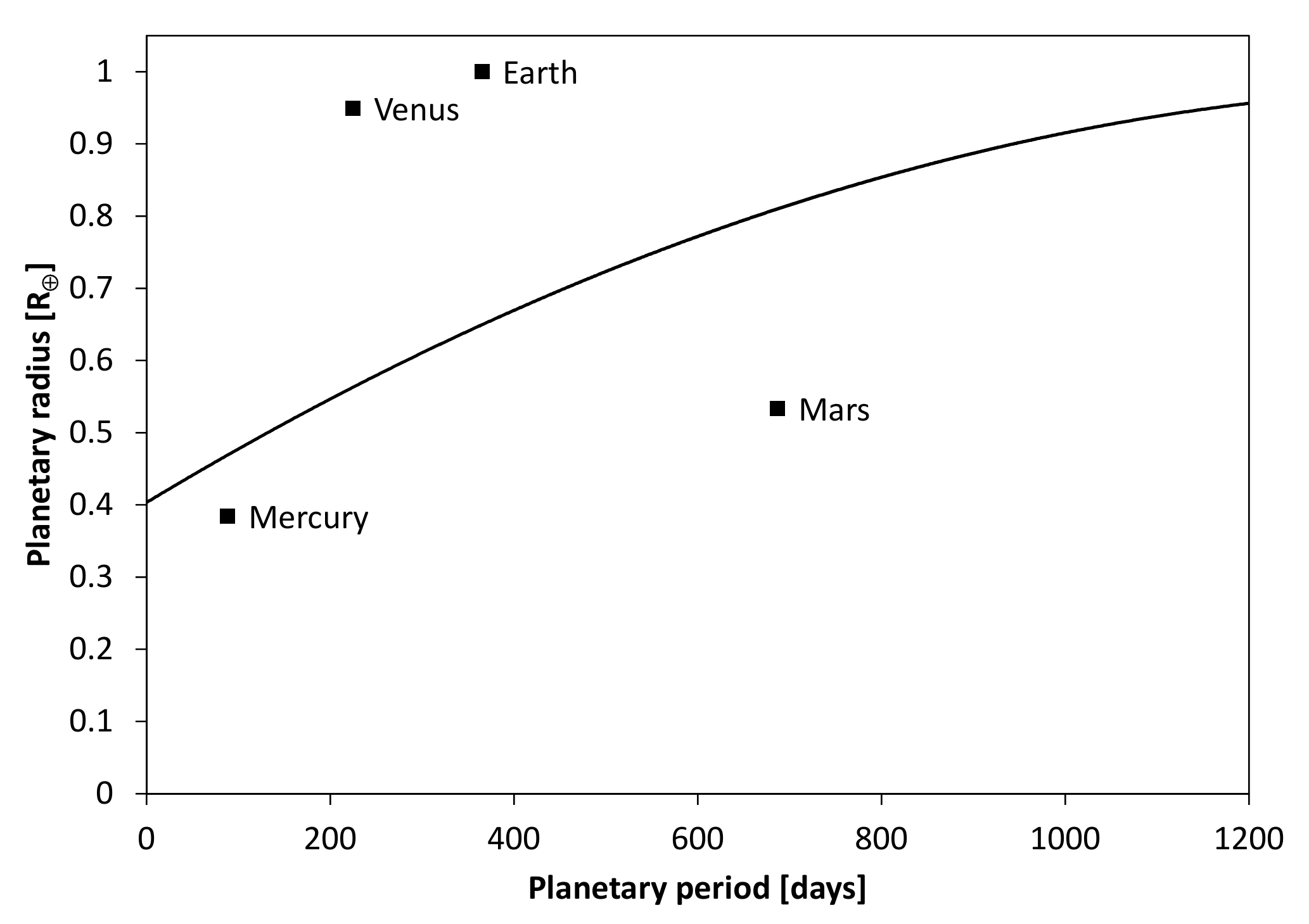}
\caption{\label{fig:snmap}Period-radius parameter space for planet transit detectability. The threshold (line) is discussed in section~\ref{sub:radiusperiod}.}
\end{figure}

\subsection{Planets interior to Mercury}
\label{sub:radiusperiod}
Our solar system does not have any planets interior to Mercury, but this is not the norm: It is estimated that $\sim$5\% of stars host (multiple) tightly-packed inner planets \citep{Lissauer2011}, and $\sim$50\% of stars have at least one $0.8-2R_{\oplus}$ planet within Mercury's orbit \citep{Fressin2013}. The latest simulations indicate that our solar system also possessed such planets, but these have been destroyed in catastrophic collisions, leaving only Mercury \citep{Volk2015}. To better characterize these configurations of solar system analogues, it would be helpful to be able to detect them in future missions. As we will see in the next section, we can expect to find most or all planets $>0.5R_{\oplus}$ within Mercury-sized orbits with \textit{PLATO 2.0}, answering interesting questions on planetary formation.

\subsection{Ultimate limits of photometry}
To determine the actual limits of detection in the radius/period parameter space, we have created a diagram of achievable S/N using a series of injections and blind retrievals. We used our Sun's data with its red noise, and injected series of transits on circular orbits with central transits, i.e. impact parameter $b=0$. Limb darkening parameters were fixed (``known'') for simplicity, as we found that their typical uncertainties account for $<1$\% of transit depth/duration variation. To check this, we have used typical uncertainties in stellar parameters and recalculated the transits with varying limb darkening coefficients from \citep{Claret2011}. For example, in the recent characterization of Kepler-138 by \citet{Jontof2015}, the stellar parameters were given as $T_{eff}=3841\pm49$K, [Fe/H]=0.280$\pm$0.099, and log(g)=4.886$\pm$0.055. With these uncertainties, the transit depth varies by $<0.5$\%.

We only varied the radius of the injected planet, and the period. Retrieval was done blindly, and considered a success if both parameters were within $10$\% of the true values.

We required the likelihood of a noise-based dip of the same depth and shape as the transits to be $<1:1,000$ in 6yrs of data, using blind recovery and our noise model, plus the \textit{PLATO 2.0} instrumental noise. With this limit, one false positive among 1,000 real planet detections is found. The measured S/N as described in section~\ref{sub:red} for such a borderline detection is $\sim$10. If the noise model is neglected, one false positive among 300 real detections occurs; to achieve the same false positive rate, the threshold would need to be increased to S/N$\sim$14. The result in Figure~\ref{fig:snmap} shows that planets $>0.5R_{\oplus}$ within Mercury-sized orbits can be detected. What is more, we can expect to detect all planets in the habitable zone of a quiet G2-dwarf, i.e. on 200--600 days periods and radii $>0.7R_{\oplus}$. This result can easily be transferred to other stellar radii. Given sufficient brightness and the same low stellar noise, the transit depth scales as $R_P^2/R_*^2$. For example, a Mars-sized planet of $0.53R_{\oplus}$ around a $0.5R_{\odot}$ M-dwarf would cause a transit dip of $\sim$97ppm, which is about the depth of Earth's transit orbiting our Sun. Consequently, and helped by shorter periods (more transits), we can expect to detect all potentially habitable planets orbiting M, K and G stars with 6yrs of \textit{PLATO 2.0} data.

It is also worth noting that the detectability function in Figure~\ref{fig:snmap} has a step at $P>1/2 T_{obs}$ (not shown in the Figure), i.e. when only one transit is observed during the mission duration. Then, all else equal, longer periods are \textit{preferred} due to their longer transit duration which makes the detection easier. For comparison, Neptune has a transit duration of 3.1d, and Jupiter ``only'' 1.4d, giving a S/N advantage of $\sqrt{3.1/1.4} \sim 1.5 \times$  from transit duration. Of course, the occurrence rate of Neptunes and Jupiters is likely different (see section~\ref{strategies}), as is their transit probability. It is important, however, that the S/N differences are taken into account when calculating population statistics with \textit{PLATO 2.0} data.

On the other side, to answer the question of what near-perfect photometry can deliver in the future, we have repeated our injections without the instrumental noise. The results for the G2 star are very similar, as most of the noise already comes from stellar trends in the \textit{PLATO 2.0} scenario. As has been shown in section~\ref{sub:earth}, the improvements towards \textit{PERFECT} are only $\sim$10\%, and up to $\sim$50\% for the smaller M-dwarfs. According to our blind retrieval simulations, even with the inclusion of standard noise models, a detection of Mercury or Mars analogues is beyond the photometric limits. This, however, is not necessarily the \textit{ultimate} limit of photometry (as asked in the introduction); we might improve noise models using proxy data from radial velocity, spectroscopy, hydrodynamical modeling, and other methods yet to be developed.

\section{Conclusion}
In this work, we have shown that future photometry will be able to detect Earth- and Venus-analogues when transiting G-dwarfs like our Sun (Table~\ref{tab:sntable}). Larger sized planets ($>2R_{\oplus}$) will be detected in a single transit around G-dwarfs, in low stellar noise cases, and assuming one can find them in the first place. The search techniques for such single transits will require further research and validation, and will likely be performed remotely, due to the large storage requirements.

While the detection of moons in a solar system configuration will remain problematic in the next decades, the situation is better for rings, and for moons in M-dwarf systems.

For source stars with strong red noise characteristics, such as our Sun, we suggest to shift the usual S/N limit from 7 to 14, in order to prevent too many false positives. This limit can be set individually for each star in question, by multiply injecting and retrieving artificial signals to the one that is in question. Alternatively, we recommend to model the noise and set an appropriate threshold for the noise model using simulations. We release all data used in this paper, plain and injected, for the community\footnote{\url{http://www.jaekle.info/injections.zip}} and encourage testing different (and also blind) retrieval techniques in preparation for \textit{PLATO 2.0} and other missions.

Despite these challenges, we believe that the era of transit planet detection is still in its infancy, and photometry will have a bright future in the coming decades.

\acknowledgements
\section*{Acknowledgements}
\begin{CJK*}{UTF8}{gkai}
We thank Jason W. Barnes and Jorge Zuluaga for their help with understanding diffractive forward-scattering during the transit of Saturn's rings, and Wei Zhu \mbox{(祝伟)} for providing data for modeling oblateness.
Daniel Angerhausen's research was supported by an appointment to the NASA Postdoctoral Program at the NASA Goddard Space Flight Center administered by Oak Ridge Associated Universities through a contract with NASA.

%\clearpage

\end{CJK*}
\end{document}